\documentclass[twocolumn]{aastex62}
\usepackage{graphicx}
\usepackage{longtable}
\usepackage{threeparttable}
\usepackage[utf8]{inputenc}
\usepackage{subfigure}
\usepackage[T1]{fontenc}

\newcommand{\Chandra}{\textit{Chandra}}
\newcommand{\Hubble}{\textit{HST}}
\newcommand{\NuSTAR}{\textit{NuSTAR}}

\begin{document}
\title{Multiwavelength Characterization of the High Mass X-ray Binary Population of M31}
\author{Margaret Lazzarini}
\author{Benjamin F. Williams}
\author{Meredith Durbin}
\author{Julianne Dalcanton}
\affiliation{Department of Astronomy, Box 351580, University of Washington, Seattle, WA 98195, USA}
\author{Vallia Antoniou}
\affiliation{Department of Physics \& Astronomy, Box 41051, Science Building, Texas Tech University, Lubbock, TX 79409, USA}
\affiliation{Center for Astrophysics | Harvard \& Smithsonian, 60 Garden Street, Cambridge, MA 02138, USA}
\author{Breanna A. Binder}
\affiliation{Department of Physics \& Astronomy, California State Polytechnic University, 3801 W. Temple Avenue, Pomona, CA 91768, USA}
\author{Michael Eracleous}
\affiliation{Department of Astronomy \& Astrophysics and Institute for Gravitation and the Cosmos, The Pennsylvania State University, 525 Davey Lab, University Par, PA 16802}
\author{Paul P. Plucinsky}
\affiliation{Center for Astrophysics | Harvard \& Smithsonian, 60 Garden Street, Cambridge, MA 02138, USA}
\author{Manami Sasaki}
\affiliation{Dr. Karl Remeis Observatory and ECAP, Universität Erlangen-Nürnberg, Sternwartstr. 7, 96049 Bamberg, Germany}
\author{Neven Vulic}
\affiliation{Laboratory for X-ray Astrophysics, NASA Goddard Space Flight Center, Code 662, Greenbelt, MD 20771, USA}
\affiliation{Department of Astronomy and Center for Space Science and Technology (CRESST), University of Maryland, College Park, MD 20742, USA}


\correspondingauthor{Margaret Lazzarini}
\email{mlazz@uw.edu}

\begin{abstract}
We present our analysis of high quality high mass X-ray binary (HMXB) candidates in M31 selected from point-source optical-counterpart candidates from the \textit{Chandra}-PHAT survey catalog. We fit the spectral energy distributions (SEDs) of optical counterpart candidates using the Bayesian Extinction and Stellar Tool (BEAST). We used the best-fit luminosity, effective temperature, radius and dust reddening for the companion stars in combination with the local star formation history, dust maps of M31, published X-ray spectral fits from \textit{XMM-Newton} observations, IR colors, and \textit{Chandra} X-ray hardness ratios to determine our best sample of HMXB candidates. The age distribution of the HMXB sample appears peaked between 10 and 50 Myr, consistent with findings in other nearby galaxies. Using the age distribution and mean SFR, we find that 80$-$136 HMXBs were produced per unit of star formation rate over the last 50 Myr and 89$-$163 HMXBs were produced per unit of star formation rate over the last 80 Myr, if we expand the assumed age limit beyond the lifetimes of single massive stars. We also calculate the HMXB production rate (HMXBs/M$_{\odot}$) over time, which ranges from $7 \times 10^{-7}$ to $4 \times 10^{-6}$ HMXBs/M$_{\odot}$ over the last 80 Myr, in agreement with both theoretical predictions and measured production rates in other galaxies.
\end{abstract}

\section{Introduction}
An X-ray binary (XRB) is a system that contains a compact object (neutron star, or black hole) that accretes mass from its stellar companion. These unique endpoints offer exceptional laboratories for testing models of massive star binary evolution. Population studies of XRBs in particular constrain models of their formation and evolution by tying their X-ray and optical properties to their parent stellar populations. Previous studies have tied the XRB population of a galaxy, in some cases including both low and high mass XRBs, to galactic properties such as the galaxy's total stellar mass \citep[e.g.,][]{Lehmer2010}, the stellar mass formed in a given star forming episode \citep[e.g.,][]{Boroson2011,Zhang2012}, the star formation rate \citep[e.g.,][]{Ranalli2003,Gilfanov2004,Mineo2012,Antoniou2010,Antoniou&Zezas2016,Lehmer2019}, and the galaxy metallicity \citep[e.g.,][]{Basuzych2013,BasuZych2016,Brorby2016}. 

XRBs for which the secondary star is also massive enough to form a neutron star or black hole are high mass X-ray binaries (HMXBs).  The HMXB phase is a key observable window in the evolution of massive binary stars, a phase of critical importance given their role as progenitors of compact object mergers detectable with gravitational waves \citep{Tauris2006}. The binary fraction of massive stars -- those with M$> 8 M_{\odot}$ that go on to form black holes and neutron stars at the end of their lives -- is at least 60\%, with the majority of these massive stellar binaries in close enough systems to interact during their lifetime \citep[e.g.;][]{Sana2012}. This high binary fraction makes it clear that understanding massive stellar evolution as a whole requires understanding massive binary stellar evolution in particular.

While relatively rare compared to their low-mass counterparts, massive stars play a major role in their host galaxies, and have long been known to ionize, chemically enrich, and inject energy into the interstellar medium \citep[e.g.;][]{Dalgarno1972,Oey1999}. More recently, however, HMXBs have emerged as another important source of feedback \citep[e.g.;][]{Garratt-Smithson2019}, and as a likely contributor to cosmic reionization \citep[e.g.;][]{Mesinger,Madau2017,Greig2018}. 

Progress in all of the above requires characterizing the HMXB production rate and understanding how the observed HMXB population depends on the age of the parent stellar population. These data, which are critical for constraining predictions of compact object merger rates \citep[e.g.;][]{Fragos2013,Andrews2018,Fragos2019}, can best be collected in nearby galaxies, where both the XRB and stellar populations can be resolved. 

Nearby galaxies are well-suited to detailed studies of XRB populations, as they can be observed in relatively few telescope pointings, and do not have the distance uncertainties and absorption issues that plague Galactic measurements. Local Group X-ray observations from \textit{Chandra}, \textit{XMM-Newton}, and \textit{NuSTAR}, combined with optical observations from the \textit{Hubble Space Telescope (HST)} have connected XRBs with their parent stellar populations \citep{Williams2018,Lazzarini2018,Garofali2018}. However, the existing data have not converged on a simple picture for the age and production rate of XRBs. For example, in the SMC, HMXBs are associated with star formation bursts between 25 and 60 Myr ago \citep{Antoniou2010,Antoniou2019}, while in the LMC they are associated with younger bursts between 6 and 25 Myr ago \citep{Antoniou&Zezas2016}. In M33, the HMXB population appears to contain an even younger component \citep[$<$5 Myr;][]{Garofali2018}. Local measurements of the HMXB production rate also span nearly an order of magnitude, ranging from 60$\pm$17 systems/($M_{\odot}$ yr$^{-1}$) in the Milky Way to 480$^{+400}_{-240}$ systems/($M_{\odot}$ yr$^{-1}$) in the low metallicity Small Magellanic Cloud \citep{Bodaghee2012,Licquia2015,Antoniou&Zezas2016,Politakis2020}, possibly indicating a trend of increasing HMXB production rate with decreasing metallicity \citep[e.g.;][]{Linden2010,Fragos2013,Ponnada2020,Fornasini2020}. Given that no clear picture has emerged from these studies, more data in different environments are needed to understand the dependence of the HMXB production rate on the properties of the parent stellar population.

An obvious next target is M31, which has both a higher metallicity and lower SFR intensity than both M33 and the Magellanic Clouds. M31 has existing high-quality optical and X-ray observations that make it possible to derive HMXB age distributions and production rates. In M31 we can pair optical observations from the Panchromatic \textit{Hubble} Andromeda Treasury (PHAT) -- a survey of one third of the star-forming disk of M31 with six band photometry for over 100 million individual stars \citep{Dalcanton,WilliamsPHAT} -- with X-ray observations from the \textit{Chandra}-PHAT survey, a \Chandra\ survey consisting of seven pointings covering most of the PHAT survey footprint \citep{Williams2018}.

We leverage several PHAT data products to investigate the M31 X-ray source population, including the 6-band catalog of resolved stellar photometry \citep{WilliamsPHAT}, the spatially resolved recent star formation history maps \citep{Lewis}, and the Bayesian Extinction and Stellar Tool (BEAST) \citep{BEAST}, a spectral energy distribution (SED) fitting tool which infers the physical parameters for optical point sources from broad-band photometry. 

The HMXB population of M31 has only recently been examined using these datasets. The main challenge has been selecting a large enough sub-sample of high quality HMXB candidates to produce a statistically robust result. Through visual inspection and spatial correlation of the \textit{Chandra}-PHAT survey, \citet{Williams2018} identified 57 X-ray sources with point source optical counterpart candidates, 8 of which were high quality HMXB candidates based on optical colors and surrounding populations. These HMXB candidates had an age distribution that peaked at 15-20 Myr and 40-50 Myr. \citet{Lazzarini2018}  analyzed the subset of \textit{Chandra}-PHAT X-ray sources that were also detected by \NuSTAR. Selecting for sources in regions with recent star formation, companion stars with SED fit parameters consistent with B-type stars, and hard X-ray colors consistent with an accreting black hole or neutron star resulted in a sample of 15 HMXB candidates, 7 of which appear in the \textit{Chandra}-PHAT ``best sample''. The \citet{Lazzarini2018} study included an additional \Chandra\ field that is not part of the \textit{Chandra}-PHAT survey, which provided additional sources. The age distribution of the \NuSTAR-selected sources broadly agreed with the \citet{Williams2018} result, with one peak at about 10 Myr and another at 25-50 Myr.

In this work, we combine likely ages from the star formation history with best-fit physical parameters for the point source X-ray optical counterpart candidates from the BEAST SED fitting tool, allowing us to select a more reliable, better characterized, and larger sample of HMXB candidates in M31. We use this sample to determine the age distribution of HMXB candidates in M31 and calculate an HMXB production rate over the last 80 Myr, because we expect HMXBs to reside in young regions. In Section \ref{data_analysis} we discuss the published \Chandra\ and \Hubble\ catalogs that were used in this study. We describe the SED fitting analysis and how we used spatially resolved star formation histories to determine the best ages for each source. We also describe how we combined the SED fits, star formation history information, and multiwavelength observations to determine our best sample of HMXB candidates. In Section \ref{results_discussion} we discuss the distribution of HMXB ages, calculate the resulting HMXB production rate in M31, and put these findings in the context of other local galaxies. We summarize our results in Section \ref{conclusions}.

We assume a distance to M31 of 776 kpc, or a distance modulus of 24.45 \citep{Dalcanton}. We assume this distance when converting from flux to luminosity and use this fixed distance in our BEAST SED fits \citep{BEAST}.

\section{Data}\label{data_analysis}
We used X-ray and optical/near-UV catalogs of the northern disk of M31 to determine the best HMXB candidate sample. The optical/near-UV source catalog is from the PHAT survey \citep{WilliamsPHAT}, and the X-ray source catalog comes from the \textit{Chandra}-PHAT survey.  The details of which, including detailed reduction techniques, counterpart candidate identification, and cross-correlations with other work, are discussed in \citep{Williams2018}. Finally, we include X-ray spectral information of the sources from the \textit{XMM-Newton} study of \citet{Sasaki2018}.

The optical, and near-UV photometry that we used in this analysis comes from the PHAT survey by \citet{Dalcanton}, which covered roughly one third of the star forming disk of M31. The final source catalog \citep{WilliamsPHAT} contains six band photometry for over 100 million individual stars. The optical and near-UV \Hubble\ bands from the PHAT survey used in this work are F814W, F475W, F336W, and F275W with central wavelengths of 8353 \AA, 4750 \AA, 3375 \AA, and 2750 \AA, respectively. We use the near-IR F110W and F160W bands (central wavelengths of 1.150 $\mu$ and 1.545 $\mu$) to screen for potential foreground stars, which fall in a very narrow sequence of F110W-F160W color (see Figure 19 from \citet{WilliamsPHAT}).

We select our sample of high mass X-ray binary candidates from the \textit{Chandra}-PHAT survey, a series of seven $\sim$50 ks \Chandra\ pointings that overlap the PHAT footprint in M31 \citep{Williams2018}. The \textit{Chandra}-PHAT survey has a limiting luminosity of $\sim 2 \times 10^{35}$ erg s$^{-1}$ in the 0.35$-$8.0 keV band and detected 373 X-ray sources in the disk of M31, 57 of which were spatially coincident with point source optical counterparts in the PHAT data. The false match probability between \textit{Chandra}-PHAT X-ray sources and O or B type stars in M31 is $\sim 2$\%, so we expect 1-2 false matches \citep{Lazzarini2018}. This false match probability was determined using the spatial density of O and B type stars in the PHAT survey photometry catalog \citep{WilliamsPHAT} and the average size of the 1$\sigma$ \Chandra\ error circles that were used to identify optical counterparts \citep{Williams2018}. We use a combination of X-ray, optical, and IR properties to rule out foreground stars and background AGN from our sample, which is described in more detail in Section~\ref{sample_selection}.

\begin{figure*}
\centering
\includegraphics[width=0.99\textwidth]{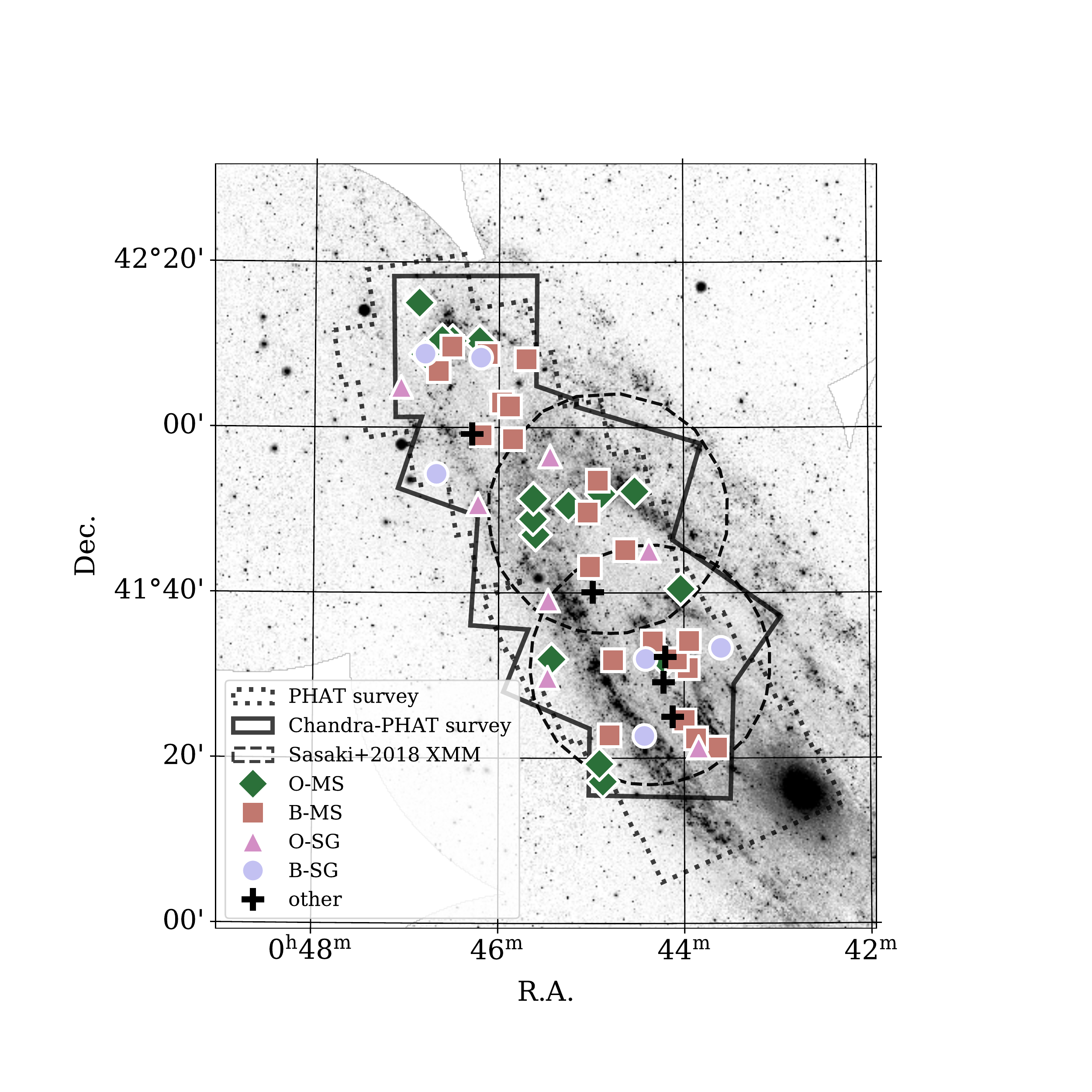}
\caption{Near UV image of M31 from the \textit{Galaxy Evolution Explorer} \citep[\textit{GALEX;}][]{GALEX} with the footprint of the Panchromatic \textit{Hubble} Andromeda Treasury survey outlined in the black dotted lines, the footprint of the Chandra-PHAT survey in the black solid lines, and the footprint of the \textit{XMM-Newton} survey by \citet{Sasaki2018} outlined in the black dashed circles. We plot the position of each of the 57 \textit{Chandra}-PHAT X-ray sources with point source optical counterparts. The color and shape of each point indicate the most likely spectral type of the companion star, based on its SED fit effective temperature, luminosity, and radius as described in Section \ref{SED}. The quality of each HMXB candidate was evaluated using a series of flags, described in Section \ref{sample_selection}. Sources plotted as black crosses do not have an effective temperature, luminosity, and radius consistent with an O or B type main sequence or supergiant star.}
\label{m31_overview}
\end{figure*}

We also include \textit{XMM-Newton} spectral fits and classifications in our analysis, which come from the \citet{Sasaki2018} survey of the northern disk of M31, which had a limiting luminosity of $\sim 7 \times 10^{34}$ erg s$^{-1}$ in the 0.5$-$2.0 keV band. \citet{Sasaki2018} classified sources using their hardness ratios and performed spectral fitting for the sources with sufficient counts ($>$100) in the 0.5$-$2.0 keV band. The \citet{Sasaki2018} catalog included 389 X-ray sources. They cross-matched sources from their catalog with the \textit{Chandra}-PHAT source catalog and found that 197 sources were detected in both surveys. The survey areas of the \textit{Chandra}-PHAT survey and the XMM-\textit{Newton} survey are similar but do not overlap perfectly. For sources in our sample with counterparts in the \citet{Sasaki2018} catalog, we list the identification number and classification from the \textit{XMM-Newton} survey in Table \ref{beast_table}.

In Figure \ref{m31_overview} we show the location of the 57 \textit{Chandra}-PHAT X-ray sources with point source optical counterpart candidates. We plot each point with a color and shape corresponding to its best-fit spectral type, as described in Section \ref{SED}. Sources plotted as black crosses do not fit the expected range of effective temperature, luminosity, and radius for O or B type main sequence or supergiant stars. We also include the outline of the PHAT survey footprint, the Chandra-PHAT survey footprint, and the \textit{XMM-Newton} survey of M31 by \citet{Sasaki2018}.

\section{Analysis}\label{analysis}
We identify the best optical counterpart candidates and best HMXB candidates using the method described briefly here and in more detail in Section~\ref{sample_selection}. We start by fitting SED models to the PHAT photometry in the UV-optical for our initial sample of 57 UV-optical counterpart candidates from \citet{Williams2018}. We then leverage the results to search for mismatches between the optical and X-ray properties that may indicate a chance superposition. We then further interpret these comparisons to better-distinguish background AGN and foreground stars from point sources more likely to lie in the disk of M31. Lastly, for the M31 sample, we combine the physical parameters inferred from the SED fits with age information from the surrounding stellar population \citep{Lewis} to identify our sample of best HMXB candidates.

\subsection{SED Fitting}\label{SED}
We fit SEDs for the identified optical counterparts to HMXB candidates using the publicly available Bayesian Extinction and Stellar Tool (BEAST) \citep{BEAST}. The BEAST fits multi-band photometry with theoretical SEDs from the Padova/PARSEC single star stellar evolution models \citep{Marigo2008,bressan2012,marigo2017}. We use four-band photometry from the PHAT survey (F275W, F336W, F475W, and F814W) and assume a fixed distance in our SED fitting as all sources are assumed to be in the disk of M31. Photometric bias and uncertainty are applied from artificial star tests performed on the data.

The BEAST fits six stellar parameters using a combination of stellar models, dust models, and models of photometric bias: age, mass, metallicity, $A_{V}$, $R_{V}$, and $f_{A}$, a parameter used to describe the mixing of different types of dust observed in the Local Group. Luminosity, effective temperature, radius, and surface gravity are then derived using the best-fit stellar models. As detailed in \citet{BEAST}, the BEAST uses a Kroupa IMF as a prior on stellar mass, a flat prior on A$_{V}$, a uniform prior on $R_{V}$ and $f_{A}$, a uniform prior on age, and a flat prior on stellar metallicity. The BEAST then then maps the mass and age priors into Hertzsprung-Russell and stellar atmosphere effective temperature versus surface gravity diagrams to produce priors on the other stellar physical parameters including luminosity, effective temperature, radius, and surface gravity.

Because the BEAST uses a probabilistic approach, any number of parameters may be fit and a probability distribution is returned for each parameter. If there are not enough data points being fit, the output distribution for each parameter will mirror the input priors. We report the parameters for the best-fit stellar models and dust models in Table \ref{beast_table}, including 16th and 84th percentile errors for each model, which reflect the probability distribution for each measurement.


To analyze the high mass stars that dominate our sample, we limited our stellar model grid to ages from log(t/yr) of 6.6 to 7.9 with steps of 0.1, matching the grid used in the SFH maps \citep{Lewis}. We limit our metallicity grid to metallicities of Z$_{initial}$=0.001, 0.004, 0.008, 0.012, 0.019, 0.03 to cover the slightly super-solar to solar range of metallicities observed in M31 \citep{Gregersen2015}. We use the default range of dust extinction values, $A_V$ of 0.0 to 10.6 magnitudes, but use the smaller step size of 0.2 mag to create a finer grid \citep{BEAST}.

\begin{figure*}
\centering
\includegraphics[width=0.99\textwidth]{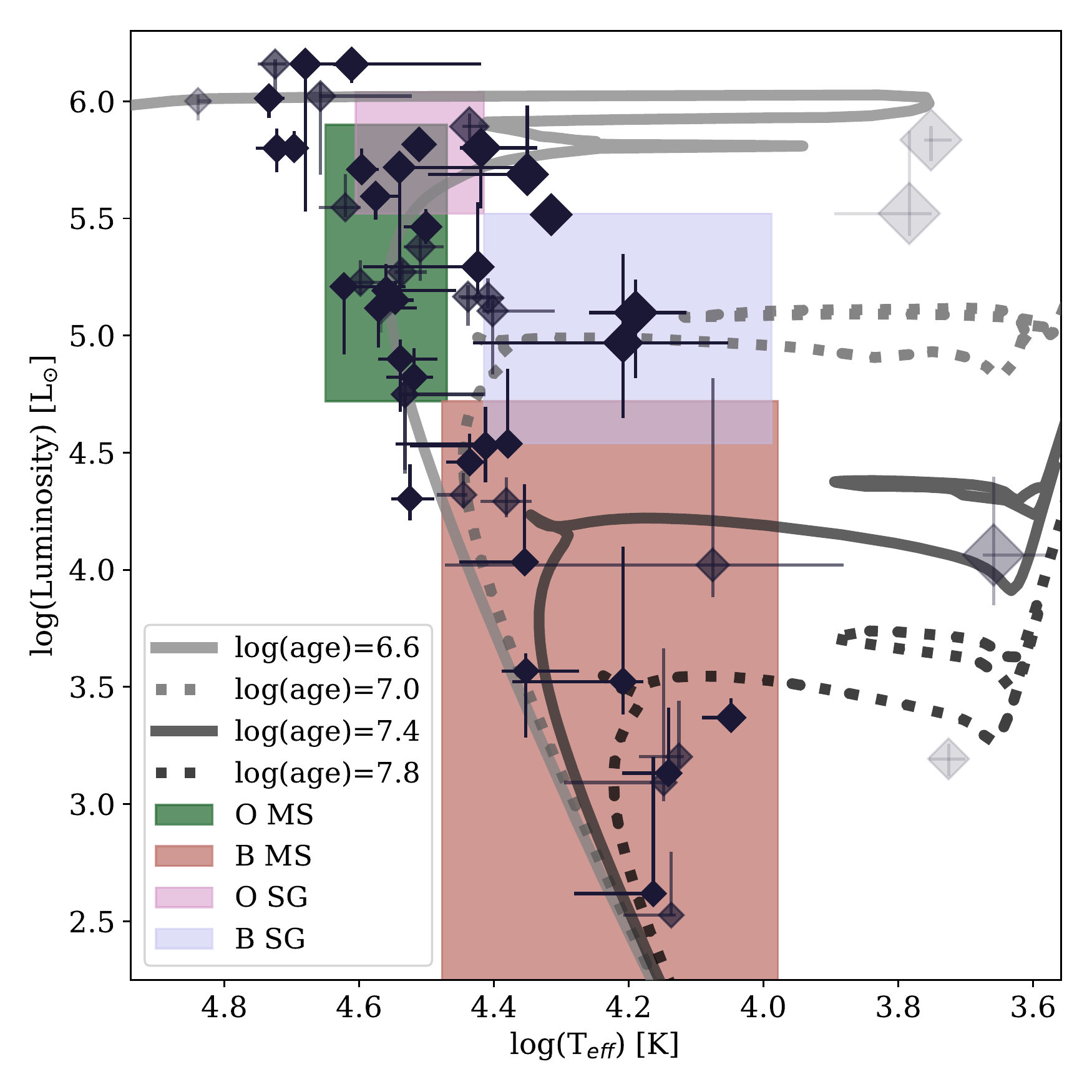}
\caption{We present the best-fit effective temperature and luminosity for the optical counterpart candidates to the \textit{Chandra}-PHAT X-ray sources with point source optical counterparts. The shaded regions represent the range of values expected for main sequence O and B type stars (O MS, B MS) and O and B type supergiants (O SG, B SG) \citep{Lamers2017}. We also plot isochrones from the Padova stellar models \citep{Marigo2008}. The size of diamond points scales with the best-fit radius for the star from its BEAST SED fit. The opacity of the diamond points depends on how many ``flags'' were raised for the source, as described in Section~\ref{sample_selection}. The darkest points represent our best sample of 33 HMXB candidates that raised zero flags and the lightest point represents the one source that raised 4 flags. See Table \ref{beast_table} for the effective temperatures, luminosities, and radii for all sources. We list the values used to evaluate the quality of each HMXB candidate in Table \ref{summary_table_values} and list the flags raised for each source in Table \ref{summary_table_flags}.}
\label{HR_digram_plot}
\end{figure*}

We use the best-fit luminosity, effective temperature, and stellar radius of the companion star from its BEAST SED fit to determine its most likely spectral type. We expect the strongest HMXB candidates to have companion stars that are consistent with the expected properties of an O or B type main sequence or supergiant star. We plot the best-fit effective temperature and luminosity for all sources in the Hertzsprung-Russell diagram in Figure \ref{HR_digram_plot}. We also include the ranges expected for O and B type main sequence and supergiant stars \citep{Lamers2017} and isochrones from the Padova stellar models at different ages \citep{Marigo2008}, at solar metallicity.

HMXB systems might host Be stars with cold disks. The H-$\alpha$ emission line does not lie within the filter set used by the PHAT survey \citep{Dalcanton}. In an emission line survey of M31 designed to overlap with the PHAT footprint by \citet{Peters2020}, Be stars were identified using narrow-band photometry. They also included the surface temperature and surface gravity for all of the B and Be stars in their catalog using the BEAST tool. When controlling for spectral type, there was no difference between the distribution of effective temperatures fit by the BEAST for the group of B and Be stars in the sample. The effective temperature of Be stars is expected to be normal compared to non-emission line B stars \citep[e.g.,][]{Porter&Rivinius2003}, so the presence of a decretion disk around the companion star should not affect our fits.

\subsection{Age Determination with Spatially Resolved Star Formation Histories}\label{sfh}
In addition to SED fitting for each individual companion star, we used spatially resolved recent star formation histories from \citet{Lewis} to determine a most likely age for each source. 

\citet{Lewis} derived the spatially resolved recent star formation history of M31 with the color magnitude diagram (CMD) fitting code, {\tt MATCH} \citep{Dolphin2002}. They divided the PHAT survey footprint into roughly nine thousand 100 by 100 pc regions (1 pc $=$ 3.76$^{\prime \prime}$ in M31). The PHAT survey is divided into 23 roughly equal sized regions, referred to as ``bricks''. The star formation history maps cover all bricks except bricks 1 and 3, which due to their location close to the galactic bulge have too high stellar density to do accurate CMD fitting. One of our HMXB candidates (004339.06$+$412117.6) falls within Brick 3 and so does not have an age determined via star formation history. 

Each 100 by 100 pc region has a measured star formation rate from log(t/yr) of 6.6 to 10.15 with a step size of 0.1. The CMD fitting optimized fits for main sequence stars, focusing the fits on the recent (<500 Myr) star formation history of M31. This range is not a limiting factor for us, as we are interested only in the star formation rate in the last 80 Myr to probe populations that could be the parent of an HMXB secondary star. The oldest of these are B-type stars, which typically have main sequence lifetimes of $<$50 Myr \citep{bressan2012}.

\begin{figure}
\centering
\includegraphics[width=0.45\textwidth]{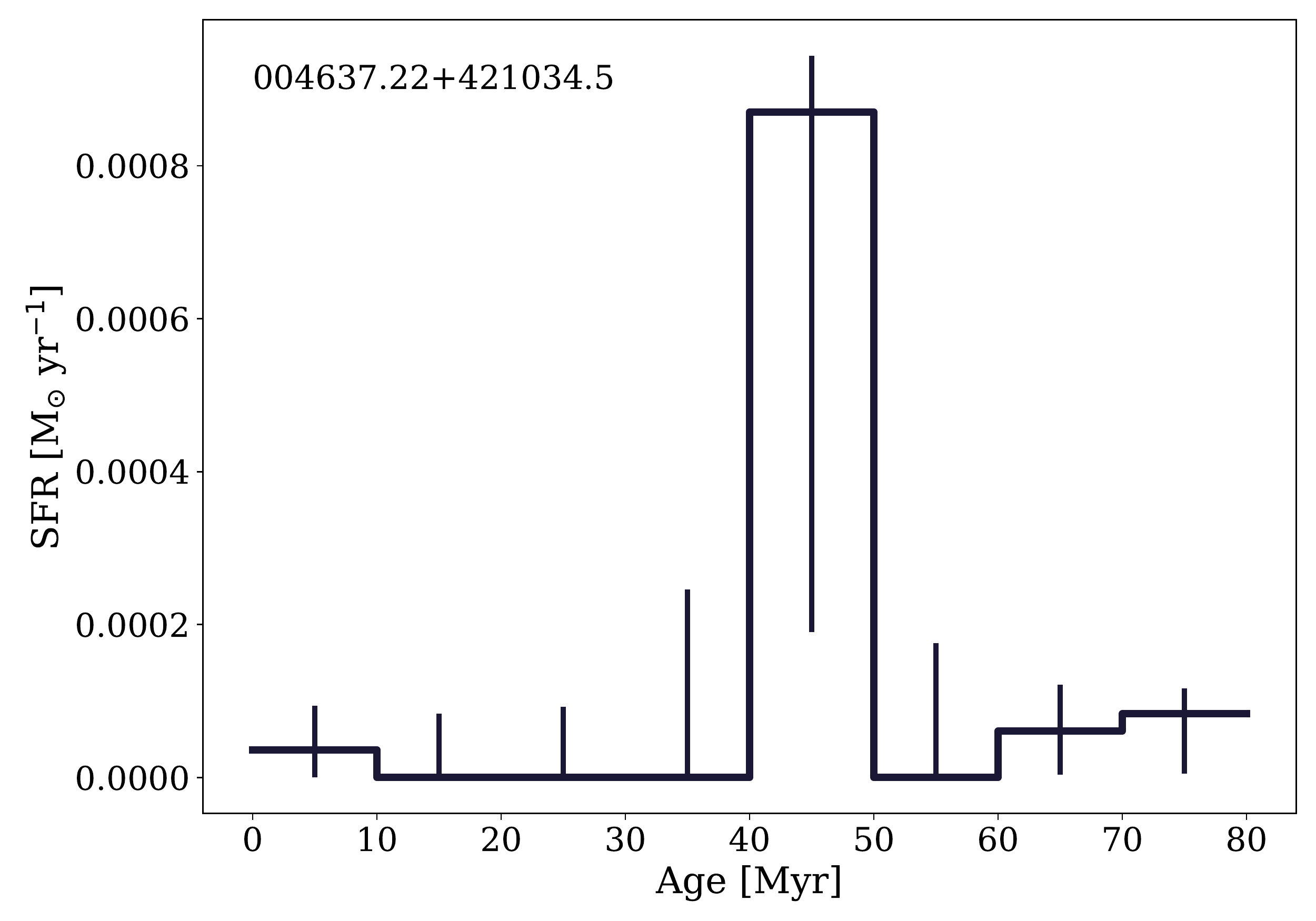}
\caption{An example of the star formation history for one 100 by 100 pc region, produced by \citet{Lewis}. This is the region surrounding the HMXB candidate in our sample, 004637.22+421034.5.}
\label{sfh_example}
\end{figure}

The errors on the star formation rate measurements were generated by the {\tt MATCH} software using hybridMC fitting. For each 100 by 100 pc region, the best fit SFR in each time bin was determined using maximum likelihood estimation. The errors represent the range of all viable star formation histories that could produce the population of stars in that region \citep{Dolphin2013}. Because the best fit star formation history often includes a best fit measurement of 0 star formation rate in some bins, the error bars are asymmetric, as the bins with 0 measured star formation rate only have upper errors. We present an example of a star formation rate measurement in the region surrounding one of our HMXB candidates, 004637.22+421034.5 in Figure \ref{sfh_example}. The SFH for this region shows a peak in the 40$-$50 Myr time bin, and there are several bins with a best fit SFR of 0.0 M$_{\odot}$ yr$^{-1}$. There is covariance in the measurements and errors for each bin, because there is a fixed amount of stellar mass in the region. The error bars are large to demonstrate this. For example, if the SFR in the 20$-$30 Myr bin was increased to the top of its upper limit, the SFR in other bins would need to be lower to conserve the total stellar mass in the region. Because the 40$-$50 Myr time bin is the only bin with significant star formation, it has a large lower error to accommodate the upper errors on all of the other bins.

We used the spatially resolved recent star formation history within the last 80 Myr in regions containing our HMXB candidates to determine their most likely ages. To do this, we calculated the total stellar mass formed in each age bin using its measured SFR and then divided this mass by the total mass formed in that region over the last 80 Myr. This measurement of the fraction of stellar mass formed in each time bin can be used as a probability distribution function (PDF), providing a measurement of the probability that the HMXB in that region has a given age. We then determine the most likely age as the age at which 50\% of the cumulative stellar mass in that region was formed, with errors showing the ages at which 16\% and 84\% of the cumulative stellar mass had formed. We list these age measurements in the column labeled `local SFH age' in Table \ref{beast_table}. This method has been used to measure the ages of HMXBs in other galaxies including the Magellanic Clouds and M33 \citep[e.g.,][]{Antoniou2010,Antoniou&Zezas2016,Williams2018,Lazzarini2018,Garofali2018,Antoniou2019}. Figure 10 in \citet{Williams2018} provides a particularly useful overview.

\subsection{Determining the Best HMXB Candidate Sample}\label{sample_selection}
We created a series of flags that were used to evaluate which HMXB candidates from the \textit{Chandra}-PHAT survey made up our best sample. We describe each flag in detail in the subsequent subsections. 

We use the measured A$_{V}$ from the PHAT dust maps \citep{Dalcanton2015}. The maps were created using near IR CMD fitting on red giant branch stars using the PHAT photometry. The maps are divided into 25 pc cells. We use the values from the map in the cell containing each HMXB candidate. The maps present the median A$_{V}$ and the width of the log-normal distribution in each 25 pc cell. We use equations in \citet{Dalcanton2015} to convert these values to a mean A$_{V}$ in each cell with the corresponding standard deviation. 

We use some of the BEAST SED fit parameters to determine the best sample. We compare the BEAST SED fit A$_{V}$ values with the measured A$_{V}$ values from the PHAT dust maps and with the \textit{XMM-Newton} N$_{H}$ measurements. We also use the most likely spectral type which we determined with the BEAST SED fit effective temperature, luminosity, and radius as described in Section \ref{SED}.

We use N$_{H}$ values from the X-ray spectral fits by \citet{Sasaki2018} using the XMM survey of M31. Sources in the \textit{XMM-Newton} catalog were matched with the \textit{Chandra}-PHAT catalog \citep{Williams2018}, and not all of the \textit{Chandra}-PHAT sources with point source optical counterpart candidates have counterparts in the \textit{XMM-Newton} survey catalog. Sources in the \textit{XMM-Newton} source catalog with more than 100 counts in the 0.2$-$12 keV band have spectral fits. There are 32 matches between our sample of 57 X-ray sources with point source optical counterparts from the \textit{Chandra}-PHAT catalog and the \textit{XMM-Newton} catalog, 21 of which had enough counts for \textit{XMM-Newton} spectral fits. 

We use the \textit{Chandra} hardness ratios for the \textit{Chandra}-PHAT X-ray sources in our analysis as well. The two hardness ratios we use are HR1: (M-S)/(H+M+S) and HR2: (H-M)/(H+M+S) where H represents the net counts in the 2$-$8 keV band, M represents the net counts in the 1$-$2 keV bad, and S represents the net counts in the 0.35$-$1 keV band.

Lastly, we used the near infrared color (F110W-F160W) of the optical counterpart candidates to flag potential foreground stars \citep{WilliamsPHAT}.

We list all of the values used to assess the quality of each HMXB candidate in Table \ref{summary_table_values}. We include a summary of which sources fit which rejection criteria, referred to as ``flags'', in Table \ref{summary_table_flags}. We describe each flag in more detail in the following sections.

Our best sample includes sources for which zero or one of the flags described below are raised. 

\subsubsection{Flag: A$_{V}$}
Our first flag evaluates whether measurements of the source's A$_{V}$ indicate that it is unlikely to be in the disk of M31, suggesting that it could be either a background galaxy or a foreground star. To raise this flag a source has to have disagreement between 1. its BEAST SED-fit A$_{V}$ and the local measured A$_{V}$ from the PHAT dust maps, 2. its BEAST SED-fit A$_{V}$ and the Hydrogen column density, N$_{H}$, measured with its \textit{XMM-Newton} spectrum, or 3. its local PHAT A$_{V}$ and the Hydrogen column density, N$_{H}$, measured with its \textit{XMM-Newton} spectrum. For a given source, if any of these three criteria are met, the A$_{V}$ flag is raised. We provide more details on how we determined whether the preceding three criteria were met in this section.

\textit{Criteria 1: }To determine whether the BEAST fit A$_{V}$ agreed with measurements of A$_{V}$ from the PHAT dust maps, we calculated the probability of measuring the BEAST SED fit A$_{V}$ given the log normal distribution of the measured A$_{V}$ from the PHAT dust maps. We include a flag value of 1 for sources with a probability of less than $6 \times 10^{-7}$, the 5$\sigma$ limit, and a flag value of 0 for sources with higher probability. The distance to the source is fixed in the BEAST SED fits. If the fit A$_{V}$ differs significantly from the A$_{V}$ value measured using CMD fitting, it could indicate that our fit makes the incorrect assumption that the source is in the disk of M31 and could be a background galaxy or foreground star, both of which would make it a poor HMXB candidate. We note excess dust reddening could be intrinsic to the HMXB system itself due to the presence of stellar winds or a decretion disk, however it is difficult to distinguish between a source that is reddened because it is far behind the disk of M31 and a source that is reddened due to expected processes in an HMXB. If a source raises this flag, it does not mean that it is not a good HMXB candidate, which is why we are using a system of flags and only ruling our sources that raise more than one flag.

\textit{Criteria 2: } There is an observed correlation between optical dust extinction, A$_{V}$ and the Hydrogen column density measured from the X-ray spectrum. \citet{Willingale2013} derived a function which relates the atomic and molecular Hydrogen column density to dust extinction, which is an improvement over previous, more simplistic, treatments of the dust-to-hydrogen ratio \citep{predehl1995,Guver2009}.

\begin{figure}
\centering
\includegraphics[width=0.45\textwidth]{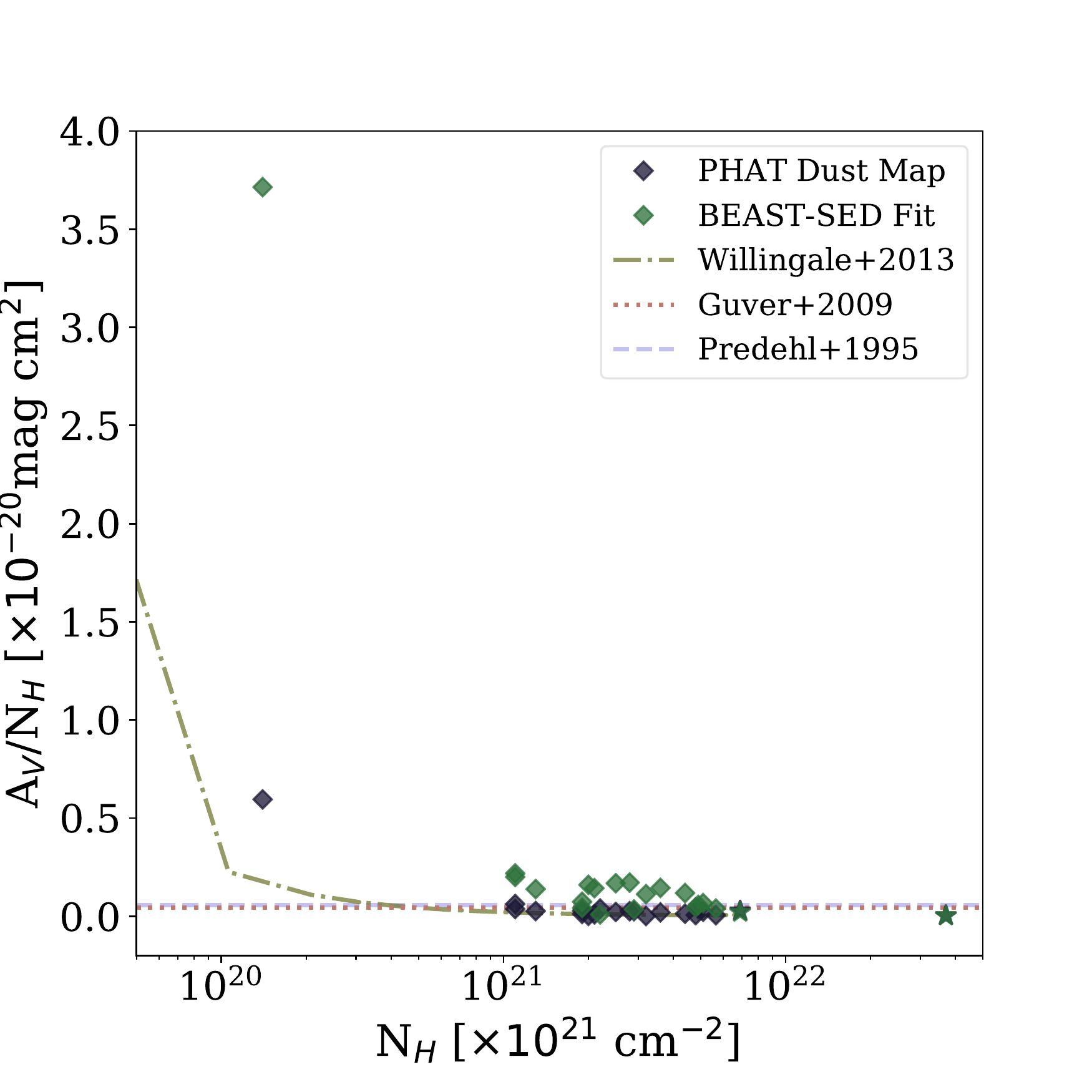}
\caption{We compare the Hydrogen column density ($N_{H}$) for X-ray sources from the \citet{Sasaki2018} \textit{XMM-Newton} catalog with the best-fit $A_{V}$ from the BEAST SED fits (lighter diamonds) and the measured $A_{V}$ from the PHAT dust maps \citep{Dalcanton2015} (darker diamonds). We plot the $N_{H}$ on the x-axis and the ratio of the $A_{V}$/$N_{H}$ on the y-axis. We include the measured relationship between A$_{V}$ and N$_{H}$ by \citet{Willingale2013}, \citet{Guver2009}, and \citet{predehl1995}. The points plotted as stars indicate sources that are suggested to be behind the disk of M31 based on their \textit{XMM-Newton} spectral fits. The inset plot is a zoomed in version of the larger plot.}
\label{nh_av}
\end{figure}

In Figure \ref{nh_av}, we plot the measured N$_{H}$ (the sum of Galactic and intrinsic components) and the ratio of A$_{V}$ to N$_{H}$ for the sources that have counterparts in the \citet{Sasaki2018} catalog with enough counts ($>$100) for a spectral fit. The \textit{XMM-Newton} spectra were fit with a power law model. Because the number of counts is low, the N$_{H}$ values should not be strongly sensitive to the details of the extinction model, making the published N$_{H}$ values suitable for our comparisons. We include lines representing the relationships measured by \citet{Willingale2013}, \citet{Guver2009}, and \citet{predehl1995}. We see a general agreement, with one obvious outlier, between the BEAST A$_{V}$ and XMM N$_{H}$ and the measured relationships.

If the A$_{V}$/N$_{H}$ value using the BEAST SED-fit A$_{V}$ plotted on the y-axis is more than a factor of two outside of the expected relationship from \citet{Willingale2013}, we call that a disagreement between the BEAST SED fit A$_{V}$ and the Hydrogen column density, N$_{H}$ and raise the A$_{V}$ flag. Only one source fails this test, 004412.04$+$413217.4, which is also an outlier compared to the PHAT A$_{V}$ (see criteria 3).

\textit{Criteria 3: } We use a similar method to set the criteria to determine whether the PHAT dust map A$_{V}$ and XMM N$_{H}$ values agree. We include the comparison between the PHAT A$_{V}$ and XMM N$_{H}$ in Figure \ref{nh_av} with the darker diamonds. The PHAT A$_{V}$ and XMM N$_{H}$ follow the expected correlation quite well, which is especially evident in the inset. 

If the ratio of the PHAT dust map A$_{V}$ and XMM N$_{H}$ differs from the expected value from \citet{Willingale2013} by more than a factor of two, as shown in Figure \ref{nh_av}, we call that a disagreement between the PHAT dust map A$_{V}$ and the \textit{XMM-Newton} N$_{H}$. Only one source raises this flag, source 004412.04$+$413217.4, which also raises the flag comparing the BEAST SED fit A$_{V}$ and \textit{XMM-Newton} N$_{H}$.

A mis-match in the BEAST SED fit A$_{V}$ or the PHAT dust map A$_{V}$ and the \textit{XMM-Newton} N$_{H}$ could indicate that the X-ray source is not in the disk of M31 and could either be a foreground star or a background galaxy.

\subsubsection{Flag: XMM Background Galaxy}
Two of the sources in our \textit{Chandra}-PHAT sample have counterparts in the \citet{Sasaki2018} \textit{XMM-Newton} sample that are classified as background galaxies based on their spectral fit. Two sources fit these criteria: 004537.67$+$415124.4 and 004502.33$+$414943.1.

While we note that these sources are classified as background galaxies based on their \textit{XMM-Newton} spectral fits, especially based on their high N$_{H}$ values, we still only rule out a source as an HMXB candidate if it fails more than one test. This is because we expect HMXBs to form in regions with high column densities, so a high measured N$_{H}$ could indicate that the HMXB is still within its natal cloud, rather than behind the disk of M31.

\subsubsection{Flag: No Spectral Type}
In Section \ref{SED} we discuss how we determined the spectral type for each source using the HR diagram presented in Figure \ref{HR_digram_plot}. 

Some sources had SED fit radii, luminosities, and effective temperatures that were not consistent with common HMXB secondary star types (O or B type main sequence or supergiant stars). These parameters suggest that their fluxes are unlikely to be dominated by a high-mass secondary star, making them less likely to be HMXBs.  We set the no spectral type flag value to 1 for these sources.

\subsubsection{Flag: Soft \textit{Chandra} Hardness Ratios}
We use the hardness ratios from the \textit{Chandra}-PHAT catalog to examine whether a source is a good HMXB candidate. HMXBs are known to have hard X-ray spectra \citep[e.g.;][]{Tullmann2011}. To meet the criteria for the soft \textit{Chandra} hardness ratios flag, a source must have HR1$<$-0.4 and HR2$<$0.1. These cutoff values were designed to flag potential foreground stars and supernova remnants using the hardness ratio diagram in Figure 8 of \citet{Williams2018} as a guide. 

Only one source, 004407.44$+$412460.0 is softer than an HMXB. We suggest that this source is a foreground star given its soft hardness ratios and point source optical counterpart.

\subsubsection{Flag: Foreground IR Colors}
Foreground stars are known to populate a relatively narrow range of F110W-F160W color space, as shown in Figure 19 of \citet{WilliamsPHAT}. On the F160W vs. F110W-F160W color magnitude diagram, foreground stars create a nearly vertical sequence between 0.4$<$F110W-F160W$<$0.8.

We raise the foreground IR colors flag for 19 sources that have 0.4$<$F110W-F160W$<$0.8. We note that a foreground IR color is not a definite sign that the optical counterpart to one of our X-ray sources is a foreground star, which is why we only remove sources from our best sample if more than one flag is raised.

\subsubsection{Best HMXB Candidate Sample Summary}
We start with the 57 sources in the \textit{Chandra}-PHAT sample of X-ray sources with point source optical counterparts. There are 33 sources which raise zero flags, 23 of which are found in regions with measured star formation in the last 80 Myr. There are 51 sources that raise one or zero flags, 35 of which are found in regions with measured star formation in the last 80 Myr. We only use sources that are found in regions with star formation in the last 80 Myr for our age analysis and HMXB production rate calculations because they have measured star formation rates in that time range. However, sources that raise one or fewer flags and are found in regions without recent star formation could still be good HMXB candidates that might be moving with a high enough velocity to have exited their birth regions.

\section{Results \& Discussion}\label{results_discussion}
We present the age distribution of a few sample selections of HMXB candidates. The ages are taken from  their local star formation histories. By combining these ages with the known total SFH of M31, we then infer the HMXB production rate for M31. HMXBs are known to exhibit significant variability, and only a fraction of systems are active at any given time. For this reason, our estimates on the production rate should be interpreted as lower limits.

\subsection{Age Distribution of HMXBs in M31}\label{age_result}
As outlined in Section \ref{sfh}, we calculated a PDF for each source, indicating the probability that the HMXB has a given age using the star formation history of the 100 pc by 100 pc region immediately surrounding it. To understand the age distribution of the full population, we summed the PDFs for all of the best HMXB candidates to produce an age distribution, the number of HMXB candidates we expect to have formed in each time bin.


We determine the errors on the age distribution by first calculating the errors on the PDF for each source. To calculate the errors on the PDFs for each source, we sampled from within the 1$\sigma$ errors on the star formation rate measurements and then generated the PDF for the region with the randomly sampled SFRs. We determined the maximum and minimum stellar mass that could be formed in each region by summing the SFR in all time bins and adding the upper and lower errors in quadrature. If a random draw of the SFR produced a total stellar mass outside these limits, we performed another random draw. We randomly sampled from the SFR errors for all HMXB candidates 10,000 times and summed the PDFs for all sources on each iteration to produce a randomly sampled overall age distribution. The errors on the age distribution represent the 16th and 84th percentile measurements.

\begin{figure*}
\centering
\includegraphics[width=0.75\textwidth]{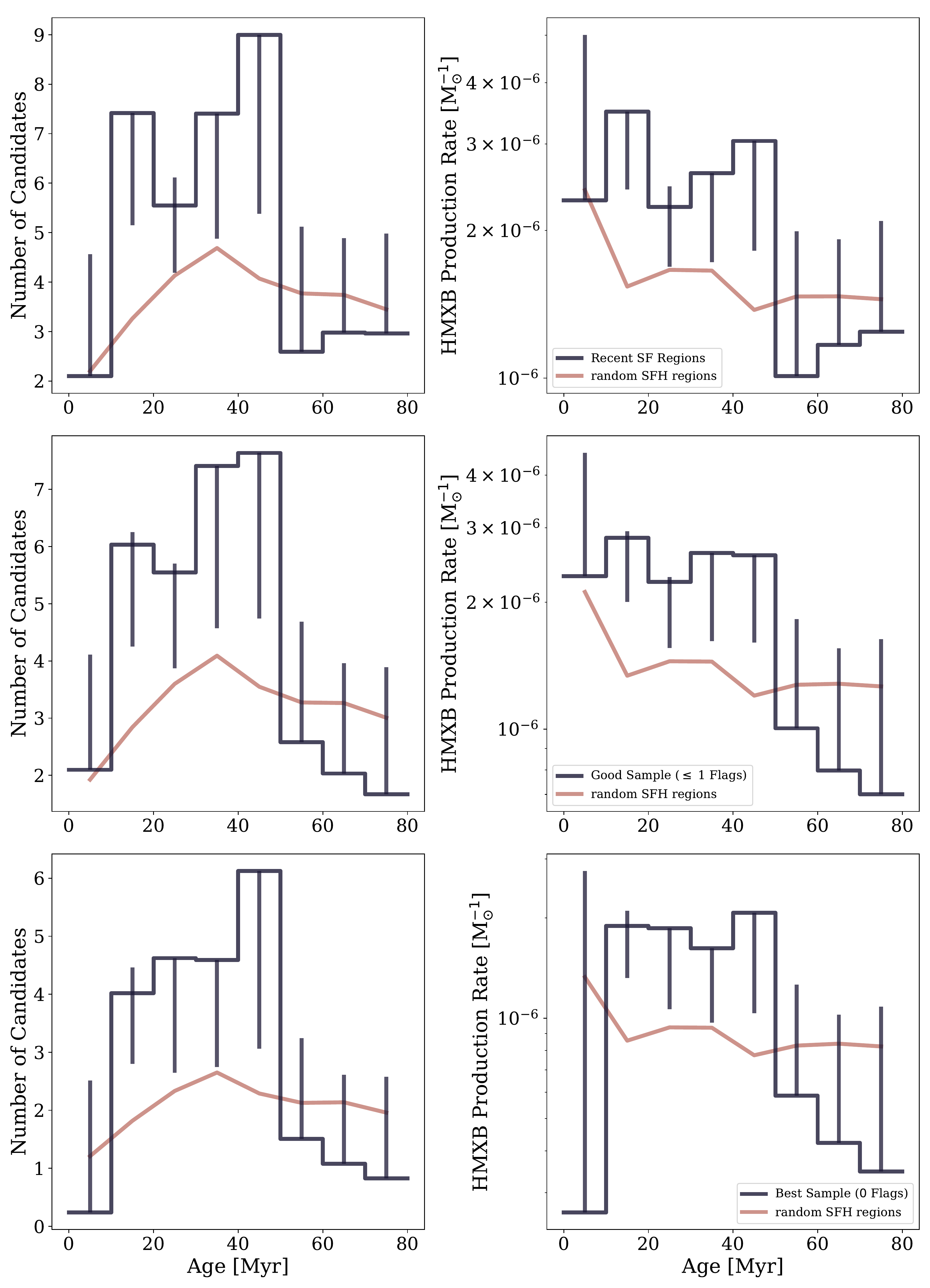}
\caption{{\scriptsize Likely age distribution and HMXB production rate using the HMXB candidates in our sample. \textbf{Top row:} The top row shows the distribution of likely ages and the production rate measurement for all of the \textit{Chandra}-PHAT sources with point source optical counterparts that are in regions with measured star formation within the last 80 Myr. The left plot shows the age distribution for our HMXB candidates. The dark histogram represents the overall age distribution for the sample, produced using the star formation history for the 100 by 100 pc region surrounding each HMXB candidate. The error bars on the histogram come from sampling within the errors on the SFR measurements 10,000 times. The pink line represents the median age distribution for an equal number of randomly selected regions from the PHAT star formation history map covered by the Chandra-PHAT survey. This random selection was performed 10,000 times. The right plot shows the HMXB production rate in units of HMXB systems produced per M$_{\odot}$. The production rate was calculated by dividing the number of candidates in each time bin by the total stellar mass formed in the PHAT survey footprint that was covered by the Chandra-PHAT survey during the same time period. The black histogram represents the HMXB candidate sample and the pink line represents the median production rate for the random sample.  \textbf{Middle row:} The middle row represents the same analysis for HMXB candidates that raised one or fewer flags as described in Section \ref{sample_selection}. This sample consists of 35 HMXB candidates. \textbf{Bottom row:} The bottom row represents the same analysis for HMXB candidates in our sample that raised zero flags, as described in Section \ref{sample_selection}. This is our best sample and comprises 23 HMXB candidates.}}
\label{pdf_prod_rate}
\end{figure*}

We present the age distributions for a few possible catalog sub-samples with the dark histograms in Figure \ref{pdf_prod_rate}, first column. The first row includes all candidates that are in regions with measured star formation within the last 80 Myr, 40 sources. The second row includes only HMXB candidates that raised one or fewer flags, as described in Section \ref{sample_selection} that reside in regions with star formation in the last 80 Myr, 35 sources. The bottom row includes only our best HMXB candidates, sources that raised zero flags and are in regions with recent star formation, totalling to 23 sources. The dark histogram represents the age distribution, with errors. 

The lighter line represents the age distribution of an equal number of randomly selected regions from within the PHAT survey area that is also covered by the Chandra-PHAT survey. In each row, we select a number of regions from the PHAT SFH map that matches the number of sources used to create the darker histogram. We calculated the PDF for each randomly selected region using its star formation rate measurements from the last 80 Myr, and then summed the PDFs from all regions to generate an overall age distribution. We performed this random selection 10,000 times and plot the median age distribution in Figure \ref{pdf_prod_rate}. Because the regions are randomly selected for each row, the shape is similar, but not identical.

We note that the error bars on our HMXB age distribution measurements are relatively large, and that some time bins only have upper errors. This is caused by the large covariance between time bins for a given SFH. Given the finite stellar mass within a region, if the best-fit SFH includes a high SFR in the 10-20 Myr bin, that will result in a lower SFR in another bin in order to conserve the total stellar mass formed. More simply, if the highest SFR allowed within errors is selected in one bin, a lower SFR within errors must be selected in another bin to maintain the same total stellar mass formed in that region in all time bins. The bins from 50 to 80 Myr show only upper errors bars, which happens when the best fit SFR is zeros, but with upper limits, so that the uncertainty only allows for the possibility of a higher mass fraction.  

The high probability portions of the histogram represent ages with high rates in their best fit, making it unlikely for a Monte Carlo selection with a higher rate in that time bin to generate an total stellar mass within the acceptable range. Thus, these distributions suggest a clear signal in the age distribution, even if there is some chance that all of the covariances and upper limits could formally conspire to reduce that signal.

We quantify the significance of our observed HMXB age distributions for each sub-sample by comparing with the age distributions produced by the 10,000 iterations where we randomly selected an equal number of regions from the PHAT star formation history maps. We wanted to know what fraction of the 10,000 random iterations for each sub-sample produced greater than or equal to the same number of candidates in the time bins between 10 and 50 Myr, where our HMXB candidate age distribution shows the largest peak. We found that zero of the 10,000 random iterations, for all of the three sub-sample sizes, produced greater than or equal to the number of candidates in our best observed HMXB age distributions for all sub-samples. This gives a $<$0.01\% probability of measuring the observed peak in our age distribution with a random sample. If we assume the lowest end of the errors on our HMXB age distributions, we found that the random draws were able to reproduce the number of candidates in the 10 to 50 Myr time range 0.23\%, 0.25\% and 0.25\% of the time, for each sub-sample respectively.

\subsection{HMXB Production Rate}\label{prod_rate_result}
We present two measures of the HMXB production rate: the number of HMXB systems formed per unit of star formation rate (HMXB systems/($M_{\odot}$ yr$^{-1}$) and the number of HMXB systems formed per solar mass (HMXB systems/$M_{\odot}$). Our HMXB production rate measurements should be interpreted as a lower limit because we only observe the fraction of HMXBs that were active at the time of observation. The observed duty cycle, or fraction of time that the XRB is active, of XRBs covers a wide range. For example, the observed duty cycle for Galactic black hole XRBs ranges from 0.2\% to 100\% with a mean value of 10\%, although these values were measured for a Galactic sample mostly consisting of low-mass XRBs \citep{Tetarenko2016}.

Single stellar evolution predicts that stars massive enough to form black holes and neutron stars should not be observed after $\sim$50 Myr. However, binary interaction may delay the onset of the supernova of the primary star because of mass transfer via binary interactions \citep[e.g.,][]{Zapartas2017}. Thus, we perform our calculation of HMXB production per unit SFR for both an assumed 50 Myr age range and an extended 80 Myr age range.

\subsubsection{Calculation of HMXBs per Unit SFR}\label{mean_prod_rate}
For direct comparison with measurements in other galaxies, we calculate the number of HMXBs produced as a function of star formation rate (HMXB systems/($M_{\odot}$ yr$^{-1}$)). To do this, we must make several assumptions related to the timescales over which HMXBs form and survive. First, we must assume a steady state, such that the global star formation rate does not change significantly over the lifetimes of massive stars ($\sim$50-80 Myr). This should be a reasonable assumption for a massive galaxy such as M31. Second, we must assume that HMXBs only form and survive over the lifetime of massive stars so that their total number is constant under the condition of a steady state star formation rate. These assumptions allow us to estimate the number of HMXBs as a function of star formation rate. 

We first find the total number of HMXB candidates in each time range (50 or 80 Myr). Our best sample, for sources that raise zero flags we find 19.6 candidates form within the last 50 Myr and 23 candidates form within the last 80 Myr. For sources that raise one or fewer flags, we find 28.7 candidates form within the last 50 Myr and 35 candidates form within the last 80 Myr. We then divide each total number of HMXB candidates by the mean total star formation rate within the PHAT footprint that is covered by the \textit{Chandra}-PHAT survey over the time range (see Section \ref{sfh}). The mean SFR over the last 50 Myr is $0.226 \pm 0.016$ M$_{\odot}$ yr$^{-1}$ and over the last 80 Myr is $0.235 \pm 0.021$ M$_{\odot}$ yr$^{-1}$.

For the sample of HMXB candidates that raised zero flags, we calculate an HMXB production rate of 89$-$107 HMXBs/(M$_{\odot}$ yr$^{-1}$) in the last 80 Myr and an HMXB production rate of 80$-$93 HMXBs/(M$_{\odot}$ yr$^{-1}$) in the last 50 Myr. For the equal sized random sample, the rate is 65$-$77 HMXBs/(M$_{\odot}$ yr$^{-1}$) in the last 80 Myr and 43$-$49 HMXBs/(M$_{\odot}$ yr$^{-1}$) in the last 50 Myr. For the sample of HMXB candidates that raised one or fewer flags, we calculate an HMXB production rate of 136$-$163 HMXBs/(M$_{\odot}$ yr$^{-1}$) in the last 80 Myr and an HMXB production rate of 118$-$136 HMXBs/(M$_{\odot}$ yr$^{-1}$) in the last 50 Myr. For the equal sized random sample, the rate is 100$-$119 HMXBs/(M$_{\odot}$ yr$^{-1}$) in the last 80 Myr and 66$-$76 HMXBs/(M$_{\odot}$ yr$^{-1}$) in the last 50 Myr.

\subsubsection{Time Resolved HMXB Production Rate}\label{time_resolved_prod_rate}
Because we have measured an age distribution as well as a time resolved recent star formation history, we are also able to calculate a time resolved HMXB production rate. We start with the summed PDF for the stellar populations immediately surrounding each HMXB candidate in the sample (shown in the left column of Figure \ref{pdf_prod_rate}), which yields the number of HMXB candidates that likely formed in each 10 Myr time bin. We then divide the PDF in each time bin by the total star formation rate within the PHAT footprint that is covered by the \textit{Chandra}-PHAT survey during that time bin. Lastly, we divide by the width of the time bin (10 Myr) to get the number of HMXB candidates we expect to form per stellar mass.

We present the HMXB production rate in each time bin in units of HMXB systems/$M_{\odot}$ in the right  column of Figure \ref{pdf_prod_rate}. The sample used to create the plots is described in Section \ref{age_result}. We also include the production rate for the random sample, shown in pink. In all three sub-samples, the HMXB production rate appears remarkably flat at early times, with a drop at 50 Myr. This relatively constant rate results in a simple conversion between these rates and our results from Section \ref{mean_prod_rate}. By multiplying the constant rate here by the characteristic time scales assumed in Section \ref{prod_rate_result} we recover the number of HMXBs formed per unit SFR. As in the age distribution plots in the left column of Figure \ref{pdf_prod_rate}, the errors on our HMXB production rates are large due to covariance between the measured SFR in each time bin.

\subsubsection{Comparison with Other Measurements}
We can compare our calculation of the HMXB production rate in units of HMXB systems/$M_{\odot}$ most directly with theoretical predictions. \citet{Linden2010} investigated how metallicity impacts the formation of bright (bolometric $L_{X}>1\times10^{36}$ erg s$^{-1}$) XRBs. At solar metallicity, they predicted an HMXB production rate of $\sim$8 HMXB systems per starburst of $10^{6}$ M$_{\odot}$, approximately 5 Myr after the starburst. Comparison between our observed population and this simulated one is difficult. The simulated population is selected using unabsorbed bolometric X-ray luminosity, and has a limiting X-ray luminosity of $\sim7\times10^{34}$ erg s$^{-1}$ in the 0.35$-$8.0 keV \textit{Chandra} band, which is lower than the limiting luminosity of $\sim2 \times 10^{35}$ erg s$^{-1}$ in the 0.35$-$8.0 keV band for the \textit{Chandra}-PHAT survey. However, our measurements agree within an order of magnitude.

We note that a direct comparison between empirical galaxy samples is difficult due to inconsistencies between how the HMXB samples were gathered in various studies. The limiting X-ray luminosity of each survey and number of X-ray observations used to gather each HMXB sample will influence the number detected because both the HMXB luminosity function and duty cycle will affect the number of sources detected. However, the X-ray luminosity function for HMXBs has a very sharp knee right around $\sim10^{35}$ erg s$^{-1}$ \citep{Sturm2013}, attributed to the propeller effect, which can inhibit accretion at low accretion rates \citep[e.g.;][]{Shtykovskiy&Gilfanov2005}. The knee in the luminosity function suggests that as long as the limiting luminosity gets down to $\sim10^{35}$ erg s$^{-1}$, the number of expected HMXBs should be comparable. Thus, as long as the duty cycles of various populations are similar, comparisons may be of interest.  

A recent observational study of the HMXB population in NGC55 \citep{Politakis2020} compared the HMXB production rate of that galaxy with those measured in other galaxies including the SMC, LMC, and Milky Way \citep{Antoniou2010,Bodaghee2012,Antoniou&Zezas2016,Antoniou2019}. In this section, we discuss the HMXB production rates of the SMC, LMC, and Milky Way that are presented in Table 5 of \citet{Politakis2020}, which were calculated with the average SFR of each galaxy. They measured an HMXB production rate in NGC55 of $299^{+50}_{-46}$ HMXBs/(M$_{\odot}$ yr$^{-1}$), down to a limiting X-ray luminosity of $3\times 10^{35}$ erg s$^{-1}$ in the \textit{Chandra} 0.3$-$7 keV energy band. The HMXB production rate has also been measured in the LMC and SMC using the average SFR for each galaxy, down to a limiting luminosity of $3\times10^{33}$ erg s$^{-1}$ in the 2$-$8 keV energy band, with production rates of $160^{+96}_{-64}$ and $480^{+400}_{-240}$ HMXBs/(M$_{\odot}$ yr$^{-1}$), respectively \citep{Antoniou&Zezas2016}. The HMXB production rate of the Milky Way has also been measured at $69 \pm 17$ HMXBs/(M$_{\odot}$ yr$^{-1}$) \citep{Bodaghee2012,Licquia2015}. The Milky Way sample was hard X-ray selected (\textit{INTEGRAL}; 20$-$40 keV) with a sensitivity limit of $3.78\times10^{-11}$ erg cm$^{-2}$ s$^{-1}$ \citep{Bodaghee2007}, which translates to a limiting luminosity of a few $\times10^{33}$ to a few $\times10^{35}$ erg s$^{-1}$, depending on the distance to the HMXB.  The agreement between our measurement of 80$-$136 HMXBs/(M$_{\odot}$ yr$^{-1}$) over the last 50 Myr and the measurement of $69 \pm 17$ HMXB systems/($M_{\odot}$ yr$^{-1}$) in the Milky Way suggests that the HMXB populations are comparable when scaled by star formation rate.

There is a known correlation between the number of HMXBs produced and the local SFR and metallicity. The number of HMXBs produced is expected to scale with the SFR \citep[e.g.;][]{Lehmer2010,Mineo2012,Antoniou2019}, a galaxy with a higher SFR is expected to produce more HMXBs. The host galaxy metallicity also plays a role. Galaxies with lower metallicity have been predicted to produce HMXBs at a higher rate due to reduced mass loss in the late stages of stellar evolution via stellar winds \citep[e.g.;][]{Linden2010,Fragos2013,Ponnada2020,Fornasini2020}. This correlation is evident in the sample of galaxies discussed in \citet{Politakis2020}, where the lowest metallicity galaxy, the SMC, has the highest HMXB production rate while the highest metallicity galaxy, the Milky Way, has the lowest HMXB production rate. We would expect M31, which has roughly solar metallicity in its disk \citep{Gregersen2015}, to have a production rate similar to that of the Milky Way, which is consistent with our measurement.

\subsection{HMXB Velocities}
Measuring the velocities of HMXB systems can be used to constrain the natal kick imparted on the compact object during its supernova. We present the methodology we used to calculate the average HMXB transverse velocity using correlations between HMXB positions and the positions of young star clusters in M31. In this section, we describe measurements made in other nearby galaxies and outline why we were unable to make this measurement for the HMXB population in M31.

This type of analysis has been done successfully in nearby galaxies and the Milky Way. Some studies distinguish the velocities of Be-XRBs (X-ray binaries with B$-$type emission line stars as the companion to the compact object), and SG-XRBs (X-ray binaries with supergiant companion stars). \citet{Antoniou&Zezas2016} found that the mean transverse velocity of HMXBs in the SMC is 13.1 km s$^{-1}$. In the LMC, they estimated a transverse velocity of $\sim12.4 \pm 7.0$ km s$^{-1}$ for HMXBs, excluding black hole or white dwarf systems. The subset of the LMC HMXB population that are confirmed NS/Be-XRB systems travel with a velocity of $\sim10.8 \pm 7.3$ km s$^{-1}$. \citet{vandenHeuvel2000} measured the transverse velocities of Be-XRBs and SG-XRBs in the Milky Way using parallax and found a mean velocity of $42\pm14$  km s$^{-1}$ for SG-XRBs and $15\pm6$  km s$^{-1}$ for Be-XRBs.

We attempted to associate our HMXB candidates with nearby young star clusters using the cluster catalog from the PHAT survey by \citep{Johnson2015} and the catalog of SED fit ages for the PHAT clusters from \citet{deMeulenaer2015}. For each HMXB candidate in our best sample, we found the distance to the nearest cluster with an age that agreed with the local stellar population within the 1$\sigma$ errors on the age of the population. The distribution of distances from the HMXB sample was indistinguishable from a sample of random positions. Thus, we cannot constrain the precise birth places of our HMXB candidates, meaning we cannot measure their velocities. It is possible that the inclination of the M31 disk leads to a young cluster density that hinders our ability to reliably match objects to their birth clusters.

\section{Conclusions}\label{conclusions}
In this work we present a study of the HMXB population of M31. We use a combination of spatially resolved star formation histories, SED fits for the optical counterparts, \Chandra\ and \textit{XMM-Newton} X-ray observations, and dust maps of M31 to characterize the largest sample of high quality HMXB candidates in M31 to date. Our main conclusions follow: 
\begin{itemize}
    \item We have presented a new method that combines SED fits of the companion stars in HMXB candidates with ages of local stellar populations from spatially resolved star formation histories, and other multiwavelength observations to select a sample of high quality HMXB candidates.
    \item We found that the dominant ages of the HMXBs in our best sample are 10$-$50 Myr. This is consistent with the studies of SMC, LMC, and M33, which show that the HMXB age depends on the star-formation history in a galaxy.
    \item We calculated that the HMXB production rate in M31 was $\sim2\times10^{-6}$ HMXB systems/M$_{\odot}$ between 10 and 50 Myr ago before dropping below $\sim1\times10^{-6}$ HMXB systems/M$_{\odot}$ after 50 Myr.
    \item We also calculate the number of HMXB systems formed over the last 50 and 80 Myr per unit SFR. For our best sample, 89-107 HMXBs/(M$_{\odot}$ yr$^{-1}$) candidates were produced over the last 80 Myr and 80-93 HMXBs/(M$_{\odot}$ yr$^{-1}$) candidates were produced over the last 50 Myr.
    \item We were not able to determine the transverse velocities of our HMXB candidates by associating them with nearby young clusters, which could be caused by the inclination of the disk of M31.
\end{itemize}

This study still marks our early efforts to understand the HMXB populations in M31. Future work will include modeling the population with binary population synthesis. For systems that are bright enough, follow-up optical spectroscopy will allow us to confirm the nature of the companion stars and time-resolved optical spectroscopy will allow us to determine the orbital parameters of any true HMXB systems.

\acknowledgements Support  for  this  work  was  provided in part by \textit{Chandra} Award Number GO5-16085X issued by the \textit{Chandra X-ray Observatory} Center, which is operated by the Smithsonian Astrophysical Observatory for and on behalf of the National Aeronautics and Space Administration under contract NAS8-03060. Support for this work was also provided in part by NASA Grant \#80NSSC19K0814. M.S. acknowledges support by the Deutsche Forschungsgemeinschaft through the Heisenberg professor grants SA 2131/5-1 and 12-1.

\facilities{Hubble Space Telescope, Chandra X-ray Observatory, XMM-Newton}

\software{astropy \citep{astropy2013,astropy2018}, matplotlib \citep{matplotlib}, SciPy \citep{scipy}, Bayesian Extinction and Stellar Tool \citep{BEAST} }

\begin{longrotatetable}
\setlength{\tabcolsep}{3pt}
\begin{deluxetable*}{llllclllllllll}
\tabletypesize{\scriptsize}
\tablecaption{SED fit parameters for point source optical counterparts to \textit{Chandra}-PHAT X-ray sources \label{beast_table}}
\tablehead{
\colhead{\textit{Chandra}-PHAT} &
\colhead{Chandra} &
\colhead{Chandra} &
\colhead{PHAT} &
\colhead{0.35$-$8.0 keV Lum.} &
\colhead{log(T$_{eff}$)} &
\colhead{log(L))} &
\colhead{R} &
\colhead{A$_{V}$} &
\colhead{M} &
\colhead{SED Age} &
\colhead{local SFH} &
\colhead{S18} &
\colhead{S18} \\
\colhead{Name} &
\colhead{R.A.} &
\colhead{Dec.} &
\colhead{Name} &
\colhead{[$\times 10^{36}$ erg s$^{-1}$]} &
\colhead{[K]} &
\colhead{[erg s$^{-1}$]} &
\colhead{[R$_{\odot}$]} &
\colhead{[mag]} &
\colhead{[M$_{\odot}$]} &
\colhead{[log(yr)]} &
\colhead{age [Myr]} &
\colhead{ID} &
\colhead{Class.}
}
\startdata
004350.76+412118.1 & 10.961532 & 41.354841 & J004350.76+412118.16 & 3.18$^{+0.28}_{-0.26}$ & 4.6$^{+0.1}_{-0.1}$ & 6.2$^{+0.1}_{-0.1}$ & 23.9$^{+2.0}_{-2.2}$ & 3.2$^{+0.2}_{-0.1}$ & 38$^{+2}_{-1}$ & 6.6$^{+0.1}_{-0.0}$ & NA & 60 & <XRB>\\
004425.73+412242.4 & 11.107230 & 41.378284 & J004425.72+412242.52 & 1.08$^{+0.17}_{-0.15}$ & 4.1$^{+0.4}_{-0.2}$ & 4.0$^{+0.8}_{-0.1}$ & 24.1$^{+8.2}_{-15.4}$ & 4.0$^{+0.1}_{-0.7}$ & 7$^{+11}_{-3}$ & 7.7$^{+0.3}_{-0.8}$ & 16$^{+47}_{-47}$ &  & \\
004452.51+411710.7 & 11.218860 & 41.286154 & J004452.52+411711.09 & 2.49$^{+0.29}_{-0.27}$ & 4.7$^{+0.1}_{-0.1}$ & 6.0$^{+0.1}_{-0.1}$ & 11.5$^{+0.9}_{-1.2}$ & 1.4$^{+0.2}_{-0.1}$ & 29$^{+3}_{-1}$ & 6.6$^{+0.1}_{-0.0}$ & NA &  & \\
004448.13+412247.9 & 11.200555 & 41.379836 & J004448.14+412248.21 & 1.31$^{+0.19}_{-0.17}$ & 4.5$^{+0.1}_{-0.1}$ & 4.9$^{+0.1}_{-0.2}$ & 7.8$^{+0.9}_{-1.2}$ & 3.6$^{+0.2}_{-0.2}$ & 19$^{+3}_{-3}$ & 6.9$^{+0.1}_{-0.3}$ & 15$^{+3}_{-3}$ & 229 & <XRB>\\
004359.83+412435.6 & 10.999298 & 41.409707 & J004359.79+412436.02 & 0.35$^{+0.15}_{-0.11}$ & 4.5$^{+0.1}_{-0.1}$ & 4.7$^{+0.1}_{-0.3}$ & 6.8$^{+1.5}_{-0.3}$ & 2.2$^{+0.2}_{-0.2}$ & 19$^{+1}_{-6}$ & 6.7$^{+0.4}_{-0.1}$ & 34$^{+4}_{-4}$ & 87 & \\
004407.44+412460.0 & 11.030994 & 41.416487 & J004407.45+412500.04 & 0.13$^{+0.07}_{-0.05}$ & 3.8$^{+0.1}_{-0.1}$ & 5.8$^{+0.1}_{-0.1}$ & 864.9$^{+109.2}_{-71.2}$ & 5.0$^{+0.2}_{-0.1}$ & 36$^{+1}_{-2}$ & 6.7$^{+0.1}_{-0.1}$ & NA & 115 & \\
004454.75+411918.3 & 11.228091 & 41.321625 & J004454.74+411917.97 & 0.89$^{+0.18}_{-0.16}$ & 4.7$^{+0.1}_{-0.1}$ & 5.8$^{+0.1}_{-0.1}$ & 10.7$^{+1.7}_{-0.4}$ & 2.8$^{+0.2}_{-0.1}$ & 22$^{+1}_{-1}$ & 6.7$^{+0.1}_{-0.1}$ & NA & 249 & <XRB>\\
004339.06+412117.6 & 10.912759 & 41.354665 & J004339.11+412117.52 & 3.54$^{+0.32}_{-0.29}$ & 4.4$^{+0.1}_{-0.1}$ & 3.6$^{+0.1}_{-0.3}$ & 4.0$^{+0.3}_{-0.4}$ & 1.0$^{+0.1}_{-0.3}$ & 8$^{+1}_{-2}$ & 7.2$^{+0.5}_{-0.2}$ & 28$^{+8}_{-8}$ & 37 & \\
004352.37+412222.8 & 10.968206 & 41.372807 & J004352.32+412223.18 & 0.31$^{+0.11}_{-0.08}$ & 4.5$^{+0.1}_{-0.1}$ & 4.7$^{+0.1}_{-0.3}$ & 6.8$^{+1.3}_{-0.4}$ & 2.0$^{+0.1}_{-0.2}$ & 19$^{+1}_{-6}$ & 6.7$^{+0.4}_{-0.1}$ & NA &  & \\
004445.88+413152.2 & 11.191097 & 41.531045 & J004445.85+413154.69 & 0.44$^{+0.17}_{-0.14}$ & 4.5$^{+0.1}_{-0.1}$ & 4.8$^{+0.1}_{-0.1}$ & 7.9$^{+1.0}_{-1.0}$ & 1.6$^{+0.2}_{-0.1}$ & 19$^{+2}_{-1}$ & 6.8$^{+0.1}_{-0.2}$ & 25$^{+3}_{-3}$ & 223 & \\
004412.17+413148.4 & 11.050810 & 41.529889 & J004412.16+413148.13 & 1.91$^{+0.23}_{-0.20}$ & 4.5$^{+0.1}_{-0.1}$ & 5.3$^{+0.1}_{-0.1}$ & 12.1$^{+1.3}_{-0.9}$ & 1.8$^{+0.2}_{-0.1}$ & 29$^{+1}_{-6}$ & 6.6$^{+0.3}_{-0.0}$ & 10$^{+65}_{-65}$ & 127 & <XRB>\\
004424.80+413201.4 & 11.103437 & 41.533501 & J004424.80+413201.30 & 3.46$^{+0.30}_{-0.27}$ & 4.2$^{+0.2}_{-0.2}$ & 5.0$^{+0.4}_{-0.3}$ & 38.9$^{+17.3}_{-17.2}$ & 4.8$^{+0.3}_{-0.4}$ & 16$^{+12}_{-4}$ & 7.1$^{+0.2}_{-0.4}$ & 15$^{+3}_{-3}$ & 157 & <XRB>\\
004357.54+413055.8 & 10.989854 & 41.515274 & J004357.53+413055.62 & 0.89$^{+0.16}_{-0.14}$ & 4.5$^{+0.1}_{-0.1}$ & 4.3$^{+0.1}_{-0.1}$ & 4.2$^{+0.8}_{-0.2}$ & 1.4$^{+0.2}_{-0.1}$ & 14$^{+2}_{-1}$ & 6.8$^{+0.2}_{-0.1}$ & 34$^{+12}_{-12}$ & 83 & <XRB>\\
004404.55+413159.4 & 11.019069 & 41.532972 & J004404.53+413159.41 & 0.28$^{+0.10}_{-0.08}$ & 4.1$^{+0.1}_{-0.1}$ & 3.1$^{+0.6}_{-0.1}$ & 5.9$^{+1.1}_{-0.7}$ & 1.2$^{+0.5}_{-0.1}$ & 5$^{+2}_{-1}$ & 7.9$^{+0.1}_{-0.4}$ & NA & 103 & <hard>\\
004420.18+413408.2 & 11.084197 & 41.568733 & J004420.17+413408.04 & 0.32$^{+0.10}_{-0.08}$ & 4.4$^{+0.1}_{-0.1}$ & 4.0$^{+0.3}_{-0.1}$ & 6.8$^{+0.6}_{-0.9}$ & 1.0$^{+0.3}_{-0.1}$ & 10$^{+3}_{-1}$ & 7.2$^{+0.1}_{-0.3}$ & 45$^{+3}_{-3}$ & 145 & <XRB>\\
004356.78+413410.9 & 10.986712 & 41.569508 & J004356.79+413410.67 & 0.30$^{+0.10}_{-0.08}$ & 4.2$^{+0.2}_{-0.1}$ & 3.5$^{+0.6}_{-0.1}$ & 7.4$^{+1.2}_{-1.5}$ & 2.2$^{+0.7}_{-0.2}$ & 7$^{+4}_{-1}$ & 7.7$^{+0.1}_{-0.5}$ & NA & 75 & <XRB>\\
004413.18+412911.4 & 11.055003 & 41.486291 & J004413.15+412910.92 & 0.65$^{+0.15}_{-0.12}$ & 3.7$^{+0.1}_{-0.1}$ & 4.1$^{+0.3}_{-0.2}$ & 172.6$^{+122.3}_{-19.5}$ & 5.2$^{+0.7}_{-1.3}$ & 8$^{+3}_{-1}$ & 7.6$^{+0.2}_{-0.2}$ & 45$^{+3}_{-3}$ & 131 & <XRB>\\
004412.04+413217.4 & 11.050227 & 41.537919 & J004412.06+413217.37 & 0.30$^{+0.18}_{-0.12}$ & 3.8$^{+0.1}_{-0.1}$ & 5.5$^{+0.4}_{-0.1}$ & 519.5$^{+69.1}_{-77.0}$ & 5.2$^{+0.8}_{-0.3}$ & 27$^{+17}_{-2}$ & 6.8$^{+0.1}_{-0.1}$ & 15$^{+3}_{-3}$ & 126 & <XRB>\\
004336.08+413320.4 & 10.900448 & 41.555509 & J004336.09+413320.61 & 0.53$^{+0.13}_{-0.11}$ & 4.4$^{+0.1}_{-0.1}$ & 5.1$^{+0.1}_{-0.3}$ & 18.7$^{+3.1}_{-1.7}$ & 2.2$^{+0.2}_{-0.1}$ & 21$^{+1}_{-5}$ & 6.9$^{+0.1}_{-0.1}$ & 30$^{+29}_{-29}$ & 33 & <XRB>\\
004402.02+414028.8 & 11.008606 & 41.674477 & J004402.06+414028.53 & 0.53$^{+0.14}_{-0.12}$ & 4.7$^{+0.1}_{-0.1}$ & 6.2$^{+0.1}_{-0.1}$ & 14.2$^{+1.9}_{-1.2}$ & 3.0$^{+0.2}_{-0.1}$ & 37$^{+2}_{-2}$ & 6.6$^{+0.1}_{-0.0}$ & NA & 93 & <XRB>\\
004528.24+412943.9 & 11.367908 & 41.495416 & J004528.24+412943.93 & 9.23$^{+0.48}_{-0.45}$ & 4.4$^{+0.1}_{-0.1}$ & 5.9$^{+0.1}_{-0.1}$ & 39.4$^{+3.2}_{-4.0}$ & 2.4$^{+0.2}_{-0.1}$ & 23$^{+1}_{-1}$ & 6.6$^{+0.1}_{-0.0}$ & 5$^{+3}_{-3}$ & 333 & <XRB>\\
004525.67+413158.2 & 11.357166 & 41.532700 & J004525.67+413158.26 & 0.26$^{+0.10}_{-0.07}$ & 4.4$^{+0.1}_{-0.1}$ & 5.2$^{+0.1}_{-0.1}$ & 16.9$^{+2.0}_{-1.3}$ & 2.8$^{+0.2}_{-0.1}$ & 23$^{+1}_{-3}$ & 6.8$^{+0.2}_{-0.1}$ & NA &  & \\
004510.96+414559.2 & 11.295728 & 41.766228 & J004510.96+414559.08 & 2.34$^{+0.25}_{-0.22}$ & 4.7$^{+0.1}_{-0.3}$ & 6.2$^{+0.1}_{-0.6}$ & 17.5$^{+12.5}_{-0.8}$ & 4.2$^{+0.5}_{-0.1}$ & 38$^{+4}_{-11}$ & 6.6$^{+0.1}_{-0.0}$ & 55$^{+7}_{-7}$ & 290 & <XRB>\\
004527.88+413905.5 & 11.366238 & 41.651368 & J004527.88+413905.55 & 2.13$^{+0.24}_{-0.22}$ & 4.4$^{+0.1}_{-0.1}$ & 5.7$^{+0.3}_{-0.1}$ & 46.3$^{+4.0}_{-12.7}$ & 5.2$^{+0.1}_{-0.2}$ & 26$^{+24}_{-1}$ & 6.6$^{+0.1}_{-0.0}$ & 39$^{+7}_{-7}$ & 331 & <XRB>\\
004459.11+414005.1 & 11.246362 & 41.667904 & J004459.09+414004.99 & 1.18$^{+0.20}_{-0.17}$ & 3.7$^{+0.1}_{-0.1}$ & 3.2$^{+0.1}_{-0.1}$ & 46.6$^{+3.6}_{-5.2}$ & 0.2$^{+0.2}_{-0.1}$ & 6$^{+1}_{-1}$ & 7.9$^{+0.0}_{-0.1}$ & 71$^{+6}_{-6}$ & 261 & <XRB>\\
004500.89+414309.8 & 11.253781 & 41.719177 & J004500.92+414309.93 & 0.70$^{+0.15}_{-0.12}$ & 4.2$^{+0.1}_{-0.1}$ & 2.6$^{+0.6}_{-0.1}$ & 3.2$^{+1.0}_{-0.3}$ & 1.2$^{+0.9}_{-0.1}$ & 4$^{+2}_{-1}$ & 7.9$^{+0.1}_{-0.5}$ & 45$^{+3}_{-3}$ & 268 & <hard>\\
004536.13+414702.5 & 11.400541 & 41.783846 & J004536.16+414702.64 & 0.23$^{+0.10}_{-0.07}$ & 4.6$^{+0.1}_{-0.1}$ & 5.7$^{+0.1}_{-0.1}$ & 15.3$^{+1.2}_{-1.7}$ & 1.4$^{+0.2}_{-0.1}$ & 44$^{+3}_{-17}$ & 6.6$^{+0.1}_{-0.0}$ & 54$^{+7}_{-7}$ &  & \\
004537.84+414856.7 & 11.407659 & 41.815539 & J004537.92+414856.53 & 0.28$^{+0.11}_{-0.08}$ & 4.6$^{+0.1}_{-0.1}$ & 5.2$^{+0.1}_{-0.1}$ & 10.0$^{+7.2}_{-0.4}$ & 1.8$^{+0.9}_{-0.1}$ & 28$^{+1}_{-4}$ & 6.6$^{+0.3}_{-0.1}$ & 45$^{+3}_{-3}$ & 360 & galaxy\\
004537.67+415124.4 & 11.406884 & 41.856688 & J004537.69+415122.88 & 0.43$^{+0.14}_{-0.12}$ & 4.6$^{+0.1}_{-0.1}$ & 5.1$^{+0.1}_{-0.1}$ & 8.7$^{+0.5}_{-1.1}$ & 1.4$^{+0.2}_{-0.1}$ & 25$^{+2}_{-1}$ & 6.7$^{+0.1}_{-0.1}$ & 64$^{+11}_{-11}$ & 358 & galaxy\\
004453.33+415159.5 & 11.222383 & 41.866435 & J004453.34+415159.45 & 0.56$^{+0.13}_{-0.11}$ & 4.5$^{+0.1}_{-0.1}$ & 5.4$^{+0.1}_{-0.1}$ & 15.6$^{+1.4}_{-2.0}$ & 3.4$^{+0.2}_{-0.1}$ & 28$^{+3}_{-2}$ & 6.8$^{+0.1}_{-0.2}$ & 34$^{+4}_{-4}$ & 247 & <XRB>\\
004455.72+415334.6 & 11.232348 & 41.892823 & J004455.72+415334.38 & 0.30$^{+0.11}_{-0.09}$ & 4.4$^{+0.1}_{-0.1}$ & 4.3$^{+0.1}_{-0.1}$ & 8.0$^{+1.3}_{-0.6}$ & 1.8$^{+0.2}_{-0.1}$ & 12$^{+1}_{-1}$ & 7.1$^{+0.2}_{-0.1}$ & 45$^{+5}_{-5}$ & 252 & <hard>\\
004422.57+414506.5 & 11.094291 & 41.751726 & J004422.58+414506.76 & 5.73$^{+0.41}_{-0.39}$ & 4.5$^{+0.1}_{-0.2}$ & 5.7$^{+0.1}_{-0.5}$ & 20.0$^{+6.9}_{-0.9}$ & 2.2$^{+0.1}_{-0.2}$ & 43$^{+1}_{-22}$ & 6.6$^{+0.3}_{-0.0}$ & 35$^{+4}_{-4}$ & 151 & <XRB>\\
004437.96+414512.6 & 11.158345 & 41.753417 & J004437.96+414512.56 & 1.33$^{+0.21}_{-0.19}$ & 4.0$^{+0.1}_{-0.1}$ & 3.4$^{+0.1}_{-0.1}$ & 12.9$^{+0.7}_{-2.4}$ & 1.4$^{+0.1}_{-0.2}$ & 6$^{+1}_{-1}$ & 7.9$^{+0.0}_{-0.1}$ & NA & 199 & <AGN>\\
004502.33+414943.1 & 11.259843 & 41.828560 & J004502.34+414943.56 & 0.33$^{+0.11}_{-0.08}$ & 4.1$^{+0.1}_{-0.1}$ & 3.2$^{+0.2}_{-0.1}$ & 7.5$^{+0.5}_{-0.9}$ & 1.0$^{+0.2}_{-0.1}$ & 5$^{+1}_{-1}$ & 7.9$^{+0.0}_{-0.1}$ & 35$^{+3}_{-3}$ & 272 & galaxy\\
004514.76+415034.5 & 11.311624 & 41.842838 & J004514.76+415034.35 & 0.73$^{+0.16}_{-0.13}$ & 4.5$^{+0.1}_{-0.1}$ & 5.2$^{+0.1}_{-0.1}$ & 10.1$^{+0.5}_{-1.4}$ & 0.8$^{+0.2}_{-0.1}$ & 26$^{+1}_{-1}$ & 6.7$^{+0.1}_{-0.1}$ & 15$^{+3}_{-3}$ & 303 & <XRB>\\
004431.82+415217.2 & 11.132787 & 41.871348 & J004431.81+415216.99 & 0.32$^{+0.11}_{-0.09}$ & 4.5$^{+0.1}_{-0.1}$ & 5.5$^{+0.1}_{-0.1}$ & 17.9$^{+1.2}_{-2.4}$ & 2.6$^{+0.2}_{-0.1}$ & 30$^{+5}_{-2}$ & 6.8$^{+0.1}_{-0.2}$ & 37$^{+23}_{-23}$ & 180 & \\
004611.38+415903.9 & 11.547594 & 41.984285 & J004611.39+415903.91 & 1.52$^{+0.22}_{-0.20}$ & 4.4$^{+0.2}_{-0.1}$ & 4.5$^{+0.3}_{-0.1}$ & 10.8$^{+0.5}_{-3.7}$ & 3.8$^{+0.3}_{-0.2}$ & 14$^{+7}_{-1}$ & 7.1$^{+0.1}_{-0.4}$ & 34$^{+7}_{-7}$ &  & \\
004617.57+415913.6 & 11.573363 & 41.986986 & J004617.57+415913.74 & 0.52$^{+0.13}_{-0.11}$ & 4.8$^{+0.0}_{-0.1}$ & 6.0$^{+0.1}_{-0.1}$ & 7.0$^{+1.3}_{-0.2}$ & 2.2$^{+0.2}_{-0.1}$ & 29$^{+1}_{-3}$ & 6.6$^{+0.1}_{-0.0}$ & 65$^{+11}_{-11}$ &  & \\
004613.49+415043.3 & 11.556365 & 41.845282 & J004613.48+415043.48 & 2.83$^{+0.32}_{-0.29}$ & 4.6$^{+0.1}_{-0.1}$ & 6.2$^{+0.1}_{-0.1}$ & 23.9$^{+0.7}_{-3.5}$ & 3.0$^{+0.2}_{-0.1}$ & 38$^{+1}_{-2}$ & 6.6$^{+0.1}_{-0.0}$ & 26$^{+36}_{-36}$ &  & \\
004640.59+415422.8 & 11.669246 & 41.906253 & J004640.58+415423.14 & 3.98$^{+0.34}_{-0.32}$ & 4.4$^{+0.1}_{-0.1}$ & 5.8$^{+0.1}_{-0.3}$ & 38.4$^{+4.4}_{-4.1}$ & 4.2$^{+0.2}_{-0.2}$ & 45$^{+1}_{-13}$ & 6.6$^{+0.1}_{-0.0}$ & NA &  & \\
004550.83+415835.1 & 11.462002 & 41.976310 & J004550.84+415835.47 & 0.40$^{+0.12}_{-0.09}$ & 4.6$^{+0.1}_{-0.1}$ & 5.2$^{+0.1}_{-0.3}$ & 7.6$^{+1.6}_{-0.7}$ & 2.6$^{+0.2}_{-0.1}$ & 28$^{+1}_{-7}$ & 6.6$^{+0.1}_{-0.0}$ & 68$^{+9}_{-9}$ &  & \\
004558.04+420302.9 & 11.492053 & 42.050671 & J004558.05+420302.99 & 2.24$^{+0.35}_{-0.31}$ & 4.6$^{+0.1}_{-0.1}$ & 5.1$^{+0.1}_{-0.2}$ & 8.7$^{+0.9}_{-0.8}$ & 1.6$^{+0.2}_{-0.1}$ & 26$^{+1}_{-5}$ & 6.6$^{+0.2}_{-0.0}$ & NA &  & \\
004543.15+415519.4 & 11.430069 & 41.922003 & J004543.17+415519.30 & 0.27$^{+0.13}_{-0.09}$ & 4.7$^{+0.1}_{-0.1}$ & 6.0$^{+0.1}_{-0.3}$ & 16.5$^{+4.4}_{-0.2}$ & 2.8$^{+0.2}_{-0.1}$ & 29$^{+14}_{-1}$ & 6.6$^{+0.1}_{-0.0}$ & 15$^{+3}_{-3}$ &  & \\
004552.94+420234.0 & 11.470802 & 42.042636 & J004552.95+420233.98 & 0.76$^{+0.16}_{-0.14}$ & 4.4$^{+0.1}_{-0.1}$ & 4.5$^{+0.1}_{-0.1}$ & 7.6$^{+1.3}_{-0.6}$ & 2.2$^{+0.2}_{-0.1}$ & 14$^{+1}_{-1}$ & 6.9$^{+0.3}_{-0.1}$ & NA &  & \\
004526.67+415631.0 & 11.361433 & 41.941835 & J004526.66+415631.40 & 1.06$^{+0.22}_{-0.19}$ & 4.4$^{+0.2}_{-0.1}$ & 5.3$^{+0.3}_{-0.1}$ & 20.9$^{+1.2}_{-8.7}$ & 3.4$^{+0.1}_{-0.6}$ & 26$^{+10}_{-3}$ & 6.8$^{+0.1}_{-0.2}$ & 45$^{+3}_{-3}$ & 326 & <hard>\\
004611.85+420827.9 & 11.549452 & 42.140872 & J004611.85+420827.88 & 5.15$^{+0.36}_{-0.33}$ & 4.2$^{+0.1}_{-0.1}$ & 5.1$^{+0.1}_{-0.3}$ & 49.2$^{+6.4}_{-4.8}$ & 4.4$^{+0.3}_{-0.1}$ & 18$^{+2}_{-5}$ & 7.0$^{+0.2}_{-0.1}$ & 27$^{+7}_{-7}$ &  & \\
004607.50+420855.7 & 11.531340 & 42.148604 & J004607.50+420855.87 & 0.68$^{+0.16}_{-0.13}$ & 4.4$^{+0.1}_{-0.1}$ & 4.5$^{+0.2}_{-0.2}$ & 9.2$^{+0.9}_{-2.9}$ & 3.8$^{+0.2}_{-0.4}$ & 14$^{+4}_{-2}$ & 7.1$^{+0.1}_{-0.3}$ & 35$^{+3}_{-3}$ &  & \\
004612.67+421027.8 & 11.552859 & 42.174164 & J004612.67+421027.80 & 0.54$^{+0.14}_{-0.11}$ & 4.6$^{+0.1}_{-0.1}$ & 5.5$^{+0.1}_{-0.1}$ & 11.3$^{+1.2}_{-1.2}$ & 3.0$^{+0.2}_{-0.1}$ & 38$^{+1}_{-12}$ & 6.6$^{+0.1}_{-0.0}$ & NA &  & \\
004630.46+421028.7 & 11.626957 & 42.174428 & J004630.47+421028.71 & 0.36$^{+0.11}_{-0.09}$ & 4.4$^{+0.1}_{-0.1}$ & 5.2$^{+0.1}_{-0.1}$ & 19.3$^{+1.9}_{-2.2}$ & 3.2$^{+0.2}_{-0.1}$ & 22$^{+2}_{-1}$ & 6.9$^{+0.1}_{-0.2}$ & 65$^{+10}_{-10}$ &  & \\
004703.82+420453.0 & 11.765938 & 42.081202 & J004703.82+420452.95 & 5.46$^{+0.40}_{-0.37}$ & 4.5$^{+0.1}_{-0.1}$ & 5.8$^{+0.1}_{-0.1}$ & 25.6$^{+2.7}_{-2.1}$ & 3.8$^{+0.2}_{-0.1}$ & 25$^{+2}_{-1}$ & 6.6$^{+0.1}_{-0.0}$ & NA &  & \\
004639.47+420649.2 & 11.664524 & 42.113468 & J004639.52+420648.81 & 0.26$^{+0.10}_{-0.08}$ & 4.1$^{+0.1}_{-0.1}$ & 2.5$^{+0.3}_{-0.1}$ & 3.2$^{+0.3}_{-0.4}$ & 0.4$^{+0.3}_{-0.1}$ & 4$^{+1}_{-1}$ & 7.9$^{+0.0}_{-0.3}$ & 55$^{+7}_{-7}$ &  & \\
004648.19+420855.4 & 11.700826 & 42.148515 & J004648.19+420855.43 & 4.14$^{+0.32}_{-0.29}$ & 4.3$^{+0.1}_{-0.1}$ & 5.5$^{+0.1}_{-0.1}$ & 44.7$^{+4.2}_{-4.3}$ & 3.6$^{+0.2}_{-0.1}$ & 31$^{+1}_{-2}$ & 6.7$^{+0.1}_{-0.1}$ & 45$^{+3}_{-3}$ &  & \\
004630.68+420947.0 & 11.627879 & 42.162824 & J004630.68+420947.03 & 0.52$^{+0.14}_{-0.11}$ & 4.1$^{+0.1}_{-0.1}$ & 3.1$^{+0.3}_{-0.1}$ & 6.4$^{+0.8}_{-0.5}$ & 3.0$^{+0.3}_{-0.1}$ & 5$^{+1}_{-1}$ & 7.9$^{+0.0}_{-0.2}$ & 29$^{+45}_{-45}$ &  & \\
004542.25+420817.9 & 11.426149 & 42.138102 & J004542.28+420818.16 & 0.83$^{+0.18}_{-0.15}$ & 4.4$^{+0.1}_{-0.1}$ & 4.3$^{+0.1}_{-0.1}$ & 6.2$^{+0.7}_{-0.7}$ & 1.8$^{+0.2}_{-0.1}$ & 13$^{+2}_{-1}$ & 7.0$^{+0.1}_{-0.3}$ & NA &  & \\
004637.22+421034.5 & 11.655142 & 42.176036 & J004637.22+421034.26 & 0.22$^{+0.11}_{-0.07}$ & 4.6$^{+0.1}_{-0.1}$ & 5.2$^{+0.1}_{-0.1}$ & 8.7$^{+1.1}_{-0.9}$ & 0.6$^{+0.2}_{-0.1}$ & 27$^{+1}_{-3}$ & 6.7$^{+0.1}_{-0.1}$ & 46$^{+4}_{-4}$ &  & \\
004652.18+421505.8 & 11.717429 & 42.251372 & J004652.05+421505.73 & 0.49$^{+0.14}_{-0.12}$ & 4.7$^{+0.1}_{-0.1}$ & 5.8$^{+0.1}_{-0.1}$ & 9.5$^{+1.5}_{-0.4}$ & 3.2$^{+0.2}_{-0.1}$ & 22$^{+4}_{-1}$ & 6.7$^{+0.1}_{-0.1}$ & NA &  & \\
004648.27+420851.1 & 11.701185 & 42.147349 & J004648.25+420851.12 & 1.49$^{+0.21}_{-0.18}$ & 4.6$^{+0.1}_{-0.1}$ & 5.6$^{+0.1}_{-0.1}$ & 14.7$^{+1.6}_{-1.2}$ & 2.8$^{+0.2}_{-0.1}$ & 39$^{+2}_{-8}$ & 6.6$^{+0.1}_{-0.0}$ & 27$^{+8}_{-8}$ &  & 
\enddata
\tablecomments{SED fit parameters for HMXB candidate companion stars. We describe each column in the order they are presented in the table. We include the name and position of each source from the \textit{Chandra}-PHAT catalog \citep{Williams2018} and the name of the HMXB candidate's companion star in the PHAT catalog \citep{WilliamsPHAT}, and its \Chandra\ 0.35$-$8.0 keV luminosity. We include the best-fit parameters for each companion star with errors from the BEAST SED fit including the effective temperature, optical luminosity, radius, dust extinction (A$_{V}$), mass, and age. We include the age of the stellar population in the 100 by 100 pc region surrounding the HMXB candidate from the \citet{Lewis} star formation history maps. Sources with a local SFH age listed as ``NA'' lie in regions without measured star formation in the last 80 Myr. For sources that were also detected in the \citet{Sasaki2018} (S18) \textit{XMM-Newton} survey of M31, we include the source ID and classification. Sources without a listed ``S18'' ID did not have a corresponding source in the S18 catalog.}
\end{deluxetable*}
\end{longrotatetable}

\begin{deluxetable}{cccc}
\tablecaption{Physical Properties Used to Classify Companion Stars\label{spectral_types}}
\tablehead{
\colhead{Stellar Type} &  
\colhead{$T_{eff}$ [K]} & 
\colhead{log(L) [$L_{\odot}$} & 
\colhead{Radius [$R_{\odot}$}
}
\startdata
O-SG & 26,000$-$40,300 & 5.52$-$6.04 & 22$-$25 \\
B-SG & 9730$-$26,000 & 4.54$-$5.52 & 25$-$66 \\
O-MS & 29,000$-$42,500 & 5.04$-$6.00 & 13$-$18 \\
B-MS & 10,100$-$29,000 & 2.02$-$5.04 & 3.4$-$13 \\
\enddata
\tablecomments{List of the ranges of effective temperatures, luminosities, and radii used to classify companion stars. For a  visual representation of these ranges, see Figure \ref{HR_digram_plot}. Values come from Appendix B of \citet{Lamers2017}.}
\end{deluxetable}

\begin{longrotatetable}
\setlength{\tabcolsep}{3pt}
\begin{deluxetable*}{lcllllccccc}

\tablecaption{Values used to determine best HMXB candidate sample \label{summary_table_values}}
\tabletypesize{\scriptsize}
\tablehead{
\colhead{\textit{Chandra}-PHAT} &
\colhead{BEAST A$_{V}$} &
\colhead{PHAT A$_{V}$} &
\colhead{A$_{V}$} &
\colhead{XMM N$_{H}$} &
\colhead{S18} &
\colhead{XMM} &
\colhead{Spectral} &
\colhead{Chandra} &
\colhead{Chandra} &
\colhead{F110W-F160W} \\
\colhead{Name} &
\colhead{[mag]} &
\colhead{[mag]} &
\colhead{PDF} &
\colhead{[cm$^{-2}$]} &
\colhead{ID} &
\colhead{Comment} &
\colhead{Type} &
\colhead{HR1} &
\colhead{HR2} &
\colhead{(IR Color)}
}
\startdata
004350.76+412118.1 & 3.20$^{+0.20}_{-0.07}$ & 0.04$\pm$1.45 & 4.63$ \times 10^{-4}$ & 2.00$ \times 10^{21}$ & 60 & XMM spec & O-SG & -0.04 & 0.63 & 0.88\\
004425.73+412242.4 & 4.00$^{+0.12}_{-0.71}$ & 1.28$\pm$0.20 & 3.28$ \times 10^{-1}$ & N/A & N/A & no XMM match & B-SG & -0.02 & 0.81 & 0.43\\
004452.51+411710.7 & 1.40$^{+0.17}_{-0.11}$ & 0.04$\pm$1.48 & 5.71$ \times 10^{-1}$ & N/A & N/A & no XMM match & O-MS & 0.03 & 0.58 & 0.94\\
004448.13+412247.9 & 3.60$^{+0.19}_{-0.24}$ & 0.04$\pm$1.43 & 2.93$ \times 10^{-5}$ & 3.20$ \times 10^{21}$ & 229 & XMM spec & B-MS & 0.12 & 0.63 & 1.40\\
004359.83+412435.6 & 2.20$^{+0.18}_{-0.17}$ & 0.06$\pm$1.30 & 3.54$ \times 10^{-2}$ & N/A & 87 & no XMM spec & B-MS & -0.10 & 0.58 & 0.61\\
004407.44+412460.0 & 5.00$^{+0.23}_{-0.09}$ & 0.31$\pm$0.69 & 1.91$ \times 10^{-5}$ & N/A & 115 & no XMM spec & none & -0.80 & -0.15 & 0.71\\
004454.75+411918.3 & 2.80$^{+0.20}_{-0.08}$ & 1.81$\pm$0.17 & 2.88$ \times 10^{-2}$ & 4.90$ \times 10^{21}$ & 249 & XMM spec & O-MS & -0.06 & 0.51 & 1.12\\
004339.06+412117.6 & 1.00$^{+0.11}_{-0.28}$ & 0.66$\pm$0.41 & 1.41$ \times 10^{-1}$ & N/A & 37 & no XMM spec & B-MS & 0.02 & 0.83 & 0.33\\
004352.37+412222.8 & 2.00$^{+0.15}_{-0.17}$ & 0.53$\pm$0.47 & 5.69$ \times 10^{-1}$ & N/A & N/A & no XMM match & B-MS & 0.26 & 0.56 & 0.60\\
004445.88+413152.2 & 1.60$^{+0.20}_{-0.10}$ & 0.94$\pm$0.35 & 3.45$ \times 10^{-1}$ & N/A & 223 & no XMM spec & B-MS & -0.37 & 0.50 & 0.17\\
004412.17+413148.4 & 1.80$^{+0.18}_{-0.10}$ & 0.34$\pm$0.69 & 4.95$ \times 10^{-1}$ & 1.30$ \times 10^{21}$ & 127 & XMM spec & O-MS & 0.00 & 0.49 & 0.54\\
004424.80+413201.4 & 4.80$^{+0.33}_{-0.40}$ & 0.72$\pm$0.41 & 5.55$ \times 10^{-3}$ & 2.80$ \times 10^{21}$ & 157 & XMM spec & B-SG & 0.04 & 0.60 & 0.84\\
004357.54+413055.8 & 1.40$^{+0.22}_{-0.09}$ & 0.23$\pm$0.84 & 8.47$ \times 10^{-1}$ & 1.90$ \times 10^{21}$ & 83 & XMM spec & B-MS & -0.07 & 0.66 & 1.15\\
004404.55+413159.4 & 1.20$^{+0.51}_{-0.03}$ & 0.38$\pm$0.62 & 9.11$ \times 10^{-1}$ & N/A & 103 & no XMM spec & B-MS & -0.36 & 0.38 & 0.60\\
004420.18+413408.2 & 1.00$^{+0.33}_{-0.02}$ & 0.76$\pm$0.33 & 3.11$ \times 10^{-2}$ & 2.90$ \times 10^{21}$ & 145 & XMM spec & B-MS & -0.33 & 0.33 & 0.22\\
004356.78+413410.9 & 2.20$^{+0.65}_{-0.21}$ & 0.28$\pm$0.74 & 1.36$ \times 10^{-1}$ & 4.80$ \times 10^{21}$ & 75 & XMM spec & B-MS & -0.05 & 0.58 & 1.01\\
004413.18+412911.4 & 5.20$^{+0.70}_{-1.26}$ & 0.71$\pm$0.38 & 1.06$ \times 10^{-3}$ & 3.60$ \times 10^{21}$ & 131 & XMM spec & none & 0.00 & 0.76 & 0.75\\
004412.04+413217.4 & 5.20$^{+0.77}_{-0.28}$ & 0.83$\pm$0.39 & 1.01$ \times 10^{-2}$ & 1.40$ \times 10^{20}$ & 126 & XMM spec & none & -0.30 & 0.11 & 0.74\\
004336.08+413320.4 & 2.20$^{+0.17}_{-0.13}$ & 0.31$\pm$0.66 & 1.38$ \times 10^{-1}$ & 5.66$ \times 10^{21}$ & 33 & XMM spec & B-SG & -0.09 & 0.66 & 0.68\\
004402.02+414028.8 & 3.00$^{+0.19}_{-0.11}$ & 0.20$\pm$0.87 & 4.33$ \times 10^{-3}$ & 2.10$ \times 10^{21}$ & 93 & XMM spec & O-MS & 0.26 & 0.54 & 0.69\\
004528.24+412943.9 & 2.40$^{+0.19}_{-0.09}$ & 0.68$\pm$0.43 & 4.04$ \times 10^{-1}$ & 1.10$ \times 10^{21}$ & 333 & XMM spec & O-SG & -0.14 & 0.51 & 0.77\\
004525.67+413158.2 & 2.80$^{+0.19}_{-0.10}$ & 0.27$\pm$0.74 & 1.41$ \times 10^{-2}$ & N/A & N/A & no XMM match & O-MS & -0.10 & 0.72 & 0.78\\
004510.96+414559.2 & 4.20$^{+0.52}_{-0.06}$ & 0.54$\pm$0.45 & 1.49$ \times 10^{-3}$ & 2.50$ \times 10^{21}$ & 290 & XMM spec & O-MS, O-SG & -0.05 & 0.64 & 1.32\\
004527.88+413905.5 & 5.20$^{+0.05}_{-0.24}$ & 0.57$\pm$0.46 & 2.52$ \times 10^{-4}$ & 4.40$ \times 10^{21}$ & 331 & XMM spec & O-SG & 0.01 & 0.77 & 1.15\\
004459.11+414005.1 & 0.20$^{+0.18}_{-0.12}$ & 0.85$\pm$0.35 & 2.36$ \times 10^{-13}$ & 2.20$ \times 10^{21}$ & 261 & XMM spec & none & -0.08 & 0.50 & 0.63\\
004500.89+414309.8 & 1.20$^{+0.91}_{-0.04}$ & 0.29$\pm$0.75 & 10.15$ \times 10^{-1}$ & N/A & 268 & no XMM spec & B-MS & -0.04 & 0.67 & 0.30\\
004536.13+414702.5 & 1.40$^{+0.17}_{-0.11}$ & 1.34$\pm$0.19 & 8.16$ \times 10^{-4}$ & N/A & N/A & no XMM match & O-MS & 0.13 & 0.78 & 0.12\\
004537.84+414856.7 & 1.80$^{+0.88}_{-0.06}$ & 1.74$\pm$0.26 & 3.65$ \times 10^{-2}$ & N/A & 360 & no XMM spec & O-MS & -0.10 & 0.79 & 0.30\\
004537.67+415124.4 & 1.40$^{+0.17}_{-0.11}$ & 1.87$\pm$0.13 & 3.28$ \times 10^{-9}$ & 6.90$ \times 10^{21}$ & 358 & galaxy & O-MS & 0.35 & 0.96 & 0.13\\
004453.33+415159.5 & 3.40$^{+0.21}_{-0.07}$ & 1.17$\pm$0.22 & 4.25$ \times 10^{-1}$ & 5.10$ \times 10^{21}$ & 247 & XMM spec & O-MS & 0.07 & 0.52 & 0.62\\
004455.72+415334.6 & 1.80$^{+0.19}_{-0.12}$ & 0.82$\pm$0.34 & 6.03$ \times 10^{-1}$ & N/A & 252 & no XMM spec & B-MS & -0.26 & 0.55 & 0.70\\
004422.57+414506.5 & 2.20$^{+0.13}_{-0.22}$ & 0.41$\pm$0.59 & 2.47$ \times 10^{-1}$ & 1.10$ \times 10^{21}$ & 151 & XMM spec & O-SG & -0.13 & 0.54 & 0.88\\
004437.96+414512.6 & 1.40$^{+0.14}_{-0.20}$ & 0.40$\pm$0.64 & 8.50$ \times 10^{-1}$ & N/A & 199 & no XMM spec & B-MS & -0.03 & 0.53 & 1.01\\
004502.33+414943.1 & 1.00$^{+0.20}_{-0.10}$ & 1.14$\pm$0.29 & 7.28$ \times 10^{-3}$ & 3.70$ \times 10^{22}$ & 272 & galaxy & B-MS & 0.02 & 0.96 & 0.18\\
004514.76+415034.5 & 0.80$^{+0.16}_{-0.12}$ & 0.54$\pm$0.55 & 1.58$ \times 10^{-1}$ & 1.90$ \times 10^{21}$ & 303 & XMM spec & O-MS & -0.05 & 0.59 & 0.00\\
004431.82+415217.2 & 2.60$^{+0.19}_{-0.09}$ & 2.23$\pm$0.12 & 1.08$ \times 10^{-5}$ & N/A & 180 & no XMM spec & O-MS & 0.01 & 0.98 & 0.26\\
004611.38+415903.9 & 3.80$^{+0.29}_{-0.25}$ & 0.80$\pm$0.35 & 5.56$ \times 10^{-2}$ & N/A & N/A & no XMM match & B-MS & 0.04 & 0.67 & 1.11\\
004617.57+415913.6 & 2.20$^{+0.19}_{-0.09}$ & 0.47$\pm$0.53 & 3.15$ \times 10^{-1}$ & N/A & N/A & no XMM match & none & -0.18 & 0.41 & 0.55\\
004613.49+415043.3 & 3.00$^{+0.20}_{-0.08}$ & 0.95$\pm$0.35 & 3.25$ \times 10^{-1}$ & N/A & N/A & no XMM match & O-SG & 0.03 & 0.73 & 1.00\\
004640.59+415422.8 & 4.20$^{+0.17}_{-0.23}$ & 0.14$\pm$1.01 & 3.47$ \times 10^{-5}$ & N/A & N/A & no XMM match & B-SG & 0.01 & 0.65 & 0.95\\
004550.83+415835.1 & 2.60$^{+0.19}_{-0.10}$ & 0.40$\pm$0.58 & 7.07$ \times 10^{-2}$ & N/A & N/A & no XMM match & B-MS & 0.08 & 0.86 & 1.12\\
004558.04+420302.9 & 1.60$^{+0.17}_{-0.11}$ & 0.53$\pm$0.47 & 8.78$ \times 10^{-1}$ & N/A & N/A & no XMM match & B-MS & -0.25 & 0.43 & 0.83\\
004543.15+415519.4 & 2.80$^{+0.20}_{-0.08}$ & 2.25$\pm$0.09 & 3.04$ \times 10^{-8}$ & N/A & N/A & no XMM match & O-MS, O-SG & 0.07 & 0.95 & 0.36\\
004552.94+420234.0 & 2.20$^{+0.18}_{-0.10}$ & 0.64$\pm$0.44 & 4.98$ \times 10^{-1}$ & N/A & N/A & no XMM match & B-MS & -0.13 & 0.70 & 0.85\\
004526.67+415631.0 & 3.40$^{+0.11}_{-0.56}$ & 1.09$\pm$0.28 & 3.21$ \times 10^{-1}$ & N/A & 326 & no XMM spec & O-SG & -0.01 & 0.35 & 1.16\\
004611.85+420827.9 & 4.40$^{+0.29}_{-0.14}$ & 0.65$\pm$0.36 & 1.19$ \times 10^{-3}$ & N/A & N/A & no XMM match & B-SG & -0.03 & 0.63 & 0.90\\
004607.50+420855.7 & 3.80$^{+0.20}_{-0.41}$ & 0.56$\pm$0.48 & 9.89$ \times 10^{-3}$ & N/A & N/A & no XMM match & B-MS & -0.21 & 0.54 & 0.82\\
004612.67+421027.8 & 3.00$^{+0.20}_{-0.08}$ & 0.85$\pm$0.29 & 2.92$ \times 10^{-1}$ & N/A & N/A & no XMM match & O-MS & 0.17 & 0.67 & 0.67\\
004630.46+421028.7 & 3.20$^{+0.20}_{-0.08}$ & 1.44$\pm$0.21 & 3.28$ \times 10^{-1}$ & N/A & N/A & no XMM match & O-MS & -0.00 & 0.48 & 0.66\\
004703.82+420453.0 & 3.80$^{+0.22}_{-0.06}$ & 0.26$\pm$0.75 & 3.16$ \times 10^{-4}$ & N/A & N/A & no XMM match & O-SG & 0.09 & 0.64 & 1.25\\
004639.47+420649.2 & 0.40$^{+0.27}_{-0.10}$ & 1.04$\pm$0.27 & 2.91$ \times 10^{-10}$ & N/A & N/A & no XMM match & B-MS & -0.11 & 0.69 & 0.21\\
004648.19+420855.4 & 3.60$^{+0.21}_{-0.07}$ & 1.18$\pm$0.29 & 2.90$ \times 10^{-1}$ & N/A & N/A & no XMM match & B-SG & 0.02 & 0.67 & 0.99\\
004630.68+420947.0 & 3.00$^{+0.26}_{-0.05}$ & 1.09$\pm$0.32 & 3.72$ \times 10^{-1}$ & N/A & N/A & no XMM match & B-MS & 0.11 & 0.19 & 1.20\\
004542.25+420817.9 & 1.80$^{+0.18}_{-0.10}$ & 0.89$\pm$0.37 & 4.90$ \times 10^{-1}$ & N/A & N/A & no XMM match & B-MS & 0.10 & 0.60 & 0.63\\
004637.22+421034.5 & 0.60$^{+0.21}_{-0.10}$ & 1.46$\pm$0.15 & 1.14$ \times 10^{-16}$ & N/A & N/A & no XMM match & O-MS & -0.16 & 0.65 & -0.00\\
004652.18+421505.8 & 3.20$^{+0.20}_{-0.07}$ & 1.59$\pm$0.20 & 2.09$ \times 10^{-1}$ & N/A & N/A & no XMM match & O-MS & -0.18 & 0.60 & 1.19\\
004648.27+420851.1 & 2.80$^{+0.20}_{-0.08}$ & 1.59$\pm$0.20 & 1.36$ \times 10^{-1}$ & N/A & N/A & no XMM match & O-MS & -0.08 & 0.54 & 1.08
\enddata
\tablecomments{Multiwavelength measurements for each source. These values were used to determine which sources from the \textit{Chandra}-PHAT catalog are the best HMXB candidates. BEAST A$_{V}$ measurements come from the BEAST SED fits. PHAT A$_{V}$ measurements come from the PHAT survey dust maps by \citet{Dalcanton2015}. The A$_{V}$ PDF value is the probability of the BEAST A$_{V}$ value given the log-normal dust distribution from the PHAT dust maps at the position of the HMXB candidate. The S18 ID and N$_{H}$ values come from the \citet{Sasaki2018} XMM survey of M31. The XMM comment column contains the following information: ``no match'' indicates that there is no match in the XMM source catalog for the \textit{Chandra}-PHAT source and the S18 ID for that source is listed as ``NA''. The XMM comment ``no spec'' means that there is a match in the XMM source catalog but there are not enough counts ($<100$) for a spectral fit. The comment ``behind M31'' indicates that the XMM spectral fit classifies the source as a galaxy behind the disk of M31. Lastly, the comment ``XMM spec'' indicates that the XMM source corresponding to the \textit{Chandra}-PHAT source has sufficient counts ($\geq100$) for a spectral fit. The Spectral Type column indicates the most likely spectral type for the HMXB candidate companion star based off of its BEAST SED fit radius, luminosity, and effective temperature. The \textit{Chandra} hardness ratios from the \textit{Chandra}-PHAT catalog are listed for each source. The hardness ratios listed are HR1=(M-S)/(H+M+S) and HR2=(H-M)/(H+M+S) where S=0.35-1.0 keV, M=1-2 keV and H=2-8 keV. The F110W-F160W (IR Color) column lists the F110W-F160W near infrared color for the optical counterpart candidate for each source from the PHAT photometry catalog.}
\end{deluxetable*}
\end{longrotatetable}

\begin{longrotatetable}
\setlength{\tabcolsep}{3pt}
\begin{deluxetable*}{ccccccc}
\tabletypesize{\scriptsize}
\tablecaption{Flags used to determine best HMXB candidate sample \label{summary_table_flags}}
\tablehead{
\colhead{\textit{Chandra}-PHAT} &
\colhead{FLAG: A$_{V}$} &
\colhead{FLAG:} &
\colhead{FLAG: No} &
\colhead{FLAG: Soft} &
\colhead{FLAG: FG} &
\colhead{FLAG} \\
\colhead{Name} &
\colhead{BEAST A$_{V}$} &
\colhead{XMM BG} &
\colhead{Spec. Type} &
\colhead{Chandra HRs} &
\colhead{IR Color} &
\colhead{SUM}
}
\startdata
004350.76+412118.1 & 0 & 0 & 0 & 0 & 0 & 0\\
004425.73+412242.4 & 0 & 0 & 0 & 0 & 1 & 1\\
004452.51+411710.7 & 0 & 0 & 0 & 0 & 0 & 0\\
004448.13+412247.9 & 0 & 0 & 0 & 0 & 0 & 0\\
004359.83+412435.6 & 0 & 0 & 0 & 0 & 1 & 1\\
004407.44+412460.0 & 0 & 0 & 1 & 1 & 1 & 3\\
004454.75+411918.3 & 0 & 0 & 0 & 0 & 0 & 0\\
004339.06+412117.6 & 0 & 0 & 0 & 0 & 0 & 0\\
004352.37+412222.8 & 0 & 0 & 0 & 0 & 1 & 1\\
004445.88+413152.2 & 0 & 0 & 0 & 0 & 0 & 0\\
004412.17+413148.4 & 0 & 0 & 0 & 0 & 1 & 1\\
004424.80+413201.4 & 0 & 0 & 0 & 0 & 0 & 0\\
004357.54+413055.8 & 0 & 0 & 0 & 0 & 0 & 0\\
004404.55+413159.4 & 0 & 0 & 0 & 0 & 1 & 1\\
004420.18+413408.2 & 0 & 0 & 0 & 0 & 0 & 0\\
004356.78+413410.9 & 0 & 0 & 0 & 0 & 0 & 0\\
004413.18+412911.4 & 0 & 0 & 1 & 0 & 1 & 2\\
004412.04+413217.4 & 1 & 0 & 1 & 0 & 1 & 3\\
004336.08+413320.4 & 0 & 0 & 0 & 0 & 1 & 1\\
004402.02+414028.8 & 0 & 0 & 0 & 0 & 1 & 1\\
004528.24+412943.9 & 0 & 0 & 0 & 0 & 1 & 1\\
004525.67+413158.2 & 0 & 0 & 0 & 0 & 1 & 1\\
004510.96+414559.2 & 0 & 0 & 0 & 0 & 0 & 0\\
004527.88+413905.5 & 0 & 0 & 0 & 0 & 0 & 0\\
004459.11+414005.1 & 1 & 0 & 1 & 0 & 1 & 3\\
004500.89+414309.8 & 0 & 0 & 0 & 0 & 0 & 0\\
004536.13+414702.5 & 0 & 0 & 0 & 0 & 0 & 0\\
004537.84+414856.7 & 0 & 0 & 0 & 0 & 0 & 0\\
004537.67+415124.4 & 1 & 1 & 0 & 0 & 0 & 2\\
004453.33+415159.5 & 0 & 0 & 0 & 0 & 1 & 1\\
004455.72+415334.6 & 0 & 0 & 0 & 0 & 1 & 1\\
004422.57+414506.5 & 0 & 0 & 0 & 0 & 0 & 0\\
004437.96+414512.6 & 0 & 0 & 0 & 0 & 0 & 0\\
004502.33+414943.1 & 0 & 1 & 0 & 0 & 0 & 1\\
004514.76+415034.5 & 0 & 0 & 0 & 0 & 0 & 0\\
004431.82+415217.2 & 0 & 0 & 0 & 0 & 0 & 0\\
004611.38+415903.9 & 0 & 0 & 0 & 0 & 0 & 0\\
004617.57+415913.6 & 0 & 0 & 1 & 0 & 1 & 2\\
004613.49+415043.3 & 0 & 0 & 0 & 0 & 0 & 0\\
004640.59+415422.8 & 0 & 0 & 0 & 0 & 0 & 0\\
004550.83+415835.1 & 0 & 0 & 0 & 0 & 0 & 0\\
004558.04+420302.9 & 0 & 0 & 0 & 0 & 0 & 0\\
004543.15+415519.4 & 1 & 0 & 0 & 0 & 0 & 1\\
004552.94+420234.0 & 0 & 0 & 0 & 0 & 0 & 0\\
004526.67+415631.0 & 0 & 0 & 0 & 0 & 0 & 0\\
004611.85+420827.9 & 0 & 0 & 0 & 0 & 0 & 0\\
004607.50+420855.7 & 0 & 0 & 0 & 0 & 0 & 0\\
004612.67+421027.8 & 0 & 0 & 0 & 0 & 1 & 1\\
004630.46+421028.7 & 0 & 0 & 0 & 0 & 1 & 1\\
004703.82+420453.0 & 0 & 0 & 0 & 0 & 0 & 0\\
004639.47+420649.2 & 1 & 0 & 0 & 0 & 0 & 1\\
004648.19+420855.4 & 0 & 0 & 0 & 0 & 0 & 0\\
004630.68+420947.0 & 0 & 0 & 0 & 0 & 0 & 0\\
004542.25+420817.9 & 0 & 0 & 0 & 0 & 1 & 1\\
004637.22+421034.5 & 1 & 0 & 0 & 0 & 0 & 1\\
004652.18+421505.8 & 0 & 0 & 0 & 0 & 0 & 0\\
004648.27+420851.1 & 0 & 0 & 0 & 0 & 0 & 0
\enddata
\tablecomments{List of flag values used to determine best HMXB candidate sample. The values used to determine the flags are listed in Table \ref{summary_table_values} and a discussion of how each cutoff was determined is in Section \ref{sample_selection}.}
\end{deluxetable*}
\end{longrotatetable}


\end{document}